\documentclass[useAMS,usenatbib]{mn2e}
\usepackage{graphicx}
\usepackage[space]{grffile}
\usepackage{latexsym}
\usepackage{amsfonts,amsmath,amssymb}
\usepackage{url}
\usepackage[utf8]{inputenc}
\usepackage{hyperref}
\hypersetup{colorlinks=false,pdfborder={0 0 0}}
\usepackage{textcomp}
\usepackage{latex_style_sf}
\usepackage{longtable}
\usepackage{multirow,booktabs}
\usepackage[english]{babel}
\usepackage{natbib}

\title[Effect of Stellar Evolution on Migrating WJs]{The Effect of Stellar Evolution on Migrating Warm Jupiters}
\author[S. Frewen \& B. Hansen] {S.~F.~N. Frewen,$^1$\thanks{E-mail: sfrewen@astro.ucla.edu} B.~M.~S. Hansen$^1$ \\
$^1$ Division of  Astronomy and Astrophysics, University of California, Los Angeles,  CA 90095-1547}

\pagerange{\pageref{firstpage}--\pageref{lastpage}} \pubyear{2015}

\begin{document}

\label{firstpage}

\maketitle



%\begin{keywords}
%planet-star interactions -- planets and satellites: dynamical evolution and stability
%\end{keywords}



\begin{abstract}
Warm jupiters are an unexpected population of extrasolar planets that are too near to their host to have formed in situ, but distant enough to retain a significant  eccentricity in the face of tidal damping. These planets are curiously absent around stars larger than two solar radii. We hypothesize that the warm jupiters are migrating due to  Kozai-Lidov oscillations, which leads to transient episodes of high eccentricity and a consequent tidal decay.
 As their host evolves, such planets would be rapidly dragged in or engulfed at minimum periapse, leading to a rapid depletion of the population with increasing stellar radius,  as is observed. Using numerical simulations, we determine the relationship between periapse distance and orbital migration rate for planets 0.1 to 10 Jupiter masses and with orbital periods between 10 and 100 days. We find that Kozai-Lidov oscillations effectively result in planetary removal early in the evolution of the host star, possibly accounting for the observed deficit. While the observed eccentricity distribution is inconsistent with the simulated distribution for an oscillating and migrating warm jupiter population, observational biases may explain the discrepancy.
\end{abstract}

\begin{keywords}
warm jupiters -- planet-star interactions -- planets and satellites: dynamical evolution and stability
\end{keywords}

\section{Introduction}\label{sec:intro}
%Background background

%background
Arguably the biggest surprise in the field of exoplanets was the discovery of hot jupiters (HJ): extrasolar planets with orbital periods less than 10 days but masses near  that of Jupiter ($M_J$) \citep{Mayor:1995vn, Butler:1997mz}. The proximity of these planets to their host precludes them from forming at their observed location \citep{Bodenheimer:2000ve}, indicating that they must have migrated after formation. A range of migration mechanisms have been proposed, including disc migration  \citep{Lin:1996ul} and planet-planet scattering \citep{Rasio:1996qf, Weidenschilling:1996pd}, but the existence of HJs that are inclined relative to the spin of their host star \citep{Hebrard:2008rr, Winn:2010bh, Triaud:2010fj, Albrecht:2012dq} indicates that at least some migrated via a mechanism that excites the planetary inclination to high values. One of these is the Kozai-Lidov (KL) mechanism \citep{Kozai:1962qy, Lidov:1962uq}, in which an inner body oscillates between highly eccentric and highly inclined modes due to an inclined, external perturber. Recent results have shown that the KL mechanism naturally leads to misaligned and flipped planetary orbits, indicating it may contribute significantly to the formation of HJs \citep{Naoz:2011yq, Naoz:2012pd, Naoz:2013vn, Li:2013qy, Teyssandier:2013fj, Petrovich:2015qe}. 

%WJs 
% Why circularization matters- mention tides?

 The efficiency of tidal circularisation is a strong function of distance, and so it is not surprising that high-eccentricity
migration is also expected to yield a population of warm jupiters (WJs).
 These planets are similar to HJs but orbit at larger periods of 10 to 100 days,  because their migration timescales are comparable to the age of the system. Many of the systems observed in this period range have  observed eccentricities too small for significant tidal evolution, but this can be understood as a consequence of the eccentricity oscillations inherent in the KL mechanism. However, the difference between a WJ population with fixed eccentricities and oscillating eccentricities becomes very important when we consider the WJ population around evolving stars.
  As stars evolve off the main sequence and increase in size, they can tidally drag in and engulf planets orbiting too closely \citep{Rasio:1996lr, Passy:2012ve, Nordhaus:2013ij, Li:2014rr}. The eccentricity of a planet plays an important role in  how long it 
survives, as  eccentricity oscillations are rapid compared to stellar evolution timescales, and so the planets will be removed when the star approaches their minimum periastron. Tidal effects are  also dramatically increased for highly eccentric planets. For planets undergoing KL oscillations, the survival of their orbits are thus determined by their maximum rather than current or observed eccentricity. An observed population (or lack thereof) of WJs around evolved stars can then give us insight into whether the population is made up of planets with constant eccentricities, or if most go through phases of significantly larger eccentricity. This process may explain the lack of HJs and WJs observed around subgiant stars \citep{Johnson:2007kx, Johnson:2010fj, Schlaufman:2013cr}, as shown in Figure 1.  % lack of jupiters around evolved stars %engulfment:  Nordhaus:2013ij they will engulf planets orbiting too close in.

In this paper we examine how a population of migrating and oscillating WJs would be affected by the evolution of their host stars compared to an observationally identical population with constant eccentricities, and determine how it compares to observations. To do so, we run numerical simulations of a WJ and a perturber over the full period range  of WJ and determine the relationship between system properties, the closest approach of the planet, and how rapidly the planets move inward. With this data  we create model populations that match the observed distribution, and calculate how such populations are winnowed by stellar evolution, both in the case where eccentricities oscillate and in control populations with constant eccentricity.
 We find that KL oscillations do cause planets to be removed much earlier in stellar evolution, in line with the observed distribution of stellar sizes for WJ hosts. The oscillations required to  produce inward migration also  skew the eccentricity distribution to values higher than those observed, but this could be due to observational biases, as we shall discuss. 

The structure of the paper is as follows: In Section \ref{sec:wjest} we estimate the number of WJs predicted around evolved stars relative to the number observed. In Section \ref{sec:dynamics} we review the relevant dynamics taking place in systems undergoing KL oscillations. In Section \ref{sec:setup} we setup our numerical simulations of migrating, oscillating WJs. In section \ref{sec:results} we discuss our numerical results, and examine the relative effect of stellar expansion on oscillating and non-oscillating  populations in Section \ref{sec:rstar}. In Section \ref{sec:observed} we compare our results to observations and discuss the possibility that observational bias explains the discrepancy between the observed and simulated eccentricity distributions. In Section \ref{sec:conc} we review our conclusions.

\begin{figure}
\begin{center}
\includegraphics[width=0.98\columnwidth]{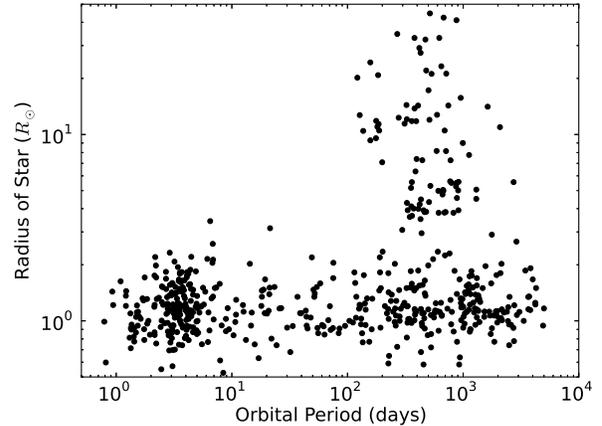}
\caption{\label{fig:PvRstar}
Confirmed exoplanets from Exoplanet Orbit Database with $M_\text{p}\sin i>0.1 M_\text{J}$ or $R_\text{p}>0.5 R_\text{J}$,  as
of June 2015. Orbital periods are shown as a function of host star radius. The WJ population is seen from 10--100 days but is
conspicuously absent around stars with radii $> 2R_{\odot}$, although more distant planets remain quite common.}
\end{center}
\end{figure}

\section{The missing warm jupiters}\label{sec:wjest}
The lack of HJs and WJs around stars larger than $2R_\odot$ is apparent from a cursory examination of Figure \ref{fig:PvRstar}. While the short-period ($\lesssim10$ days) orbits of HJs naturally lead to their engulfment early on in stellar evolution, the lack of WJs at similar stellar sizes in spite of orbits $\sim10$ times larger is surprising. However, the number of WJs around stars of all sizes, and the number of exoplanets around evolved stars at all periods, are both significantly lower than in other regions of exoplanet period--stellar-radius parameter space. This contrast raises the question of whether the number of WJs around larger stars is genuinely below observational predictions, or if they simply suffer from poor statistics. We answer this question using the observed number of WJs around main-sequence stars ($R_*=1-2 R_\odot$) and their observed periapse values, combined with the observed number of lukewarm Jupiters (LJs, periods of $100-1000$ days). 

The number of planets observed around stars of a radius $R_*$ is
\begin{align}\label{eq:nobs}
N_\text{p,obs}(R_*) = N_{*,\text{obs}}(R_*)f_\text{p,0}f_\text{p,S}(R_*)
\end{align}
where  $N_{*,\text{obs}}(R_*)$ is the number of stars observed at a given stellar radius, $f_\text{p,0}$ is the initial frequency of planets around the stars, and $f_\text{p,S}(R_*)$ is the fraction of planets that have survived prior stellar evolution.

For WJs and LJs around main-sequence stars, we get
\begin{align}
N_\text{WJ}(1-2R_\odot) &=  N_{*,\text{obs}}(1-2R_\odot)f_\text{WJ,0}f_\text{WJ,S}(1-2R_\odot)\\
N_\text{LJ}(1-2R_\odot) &=  N_{*,\text{obs}}(1-2R_\odot)f_\text{LJ,0}f_\text{LJ,S}(1-2R_\odot)
\end{align}

Assuming survival rates are similar without stellar evolution, the relative fraction of stars with WJs and LJs is:
\begin{align}
\frac{f_\text{WJ,0}}{f_\text{LJ,0}} = \frac{N_\text{WJ}(1-2R_\odot)}{N_\text{LJ}(1-2R_\odot)}
\end{align}

Equation \ref{eq:nobs} holds for larger stellar radii as well:
\begin{align}
N_\text{WJ}(>2R_\odot) &=  N_{*,\text{obs}}(>2R_\odot)f_\text{WJ,0}f_\text{WJ,S}(>2R_\odot)\\
N_\text{LJ}(>2R_\odot) &=  N_{*,\text{obs}}(>2R_\odot)f_\text{LJ,0}f_\text{LJ,S}(>2R_\odot)
\end{align}

Dividing these two equations, we get the predicted number of WJs around evolved stars based on their observed number around main-sequence stars and observed number of HJs:
\begin{align}
N_\text{WJ}(>2R_\odot) &=  \frac{N_\text{WJ}(1-2R_\odot)}{N_\text{LJ}(1-2R_\odot)}\frac{f_\text{WJ,S}(>2R_\odot)}{f_\text{LJ,S}(>2R_\odot)}N_\text{LJ}(>2R_\odot)
\end{align}

Assuming all LJs survive ($f_\text{LJ,S}(>2R_\odot)=1$) gives the most conservative estimate for the number of WJs. Observations provide values for $N_\text{WJ}(1-2R_\odot)$, $N_\text{LJ}(1-2R_\odot)$, and $N_\text{LJ}(R_*>2R_\odot)$, as detailed below, so an estimate for $f_\text{WJ,S}(>2R_\odot)$ allowed us to calculate the predicted number of WJs around evolved stars, $N_\text{WJ}(R_*>2R_\odot)$. To do so we assumed the observed periapse distribution for WJs around unevolved stars is representative of the true periapse distribution. We also assumed that exoplanets are removed when their periapse comes within 2.5 stellar radii of their host (based on the smallest observed periapse-to-stellar-radius ratio, 2.7, in the case of WASP-12b as reported by \citealt{Maciejewski:2011fv}). 

The data for this estimation came from Exoplanet Orbit Database \citep{Wright:2011fk}. We limited our dataset to massive planets\footnote{Due to some anomalously low values in the \texttt{MASS} keyword, we included planets with either \texttt{MSINI > 0.1} or \texttt{R > 0.5}.} with listed eccentricity\footnote{To avoid excluding planets on circular orbits, we used the filter \texttt{ECC > -1}.} values. We also excluded possible brown dwarfs\footnote{using the limit \texttt{MASS < 10}.}, as such massive bodies may have formed via a different mechanism than exoplanets. Finally, we used the periapse distribution for planets around stars with radii $1-2R_\odot$, rather than including planets around smaller or larger stars\footnote{The periapse dataset used \texttt{RSTAR >= 1.0 and RSTAR < 2.0}.}. From these values we calculated the predicted number of observed WJs as a function of stellar radius. 

\begin{figure}
\begin{center}
\includegraphics[width=1.0\columnwidth]{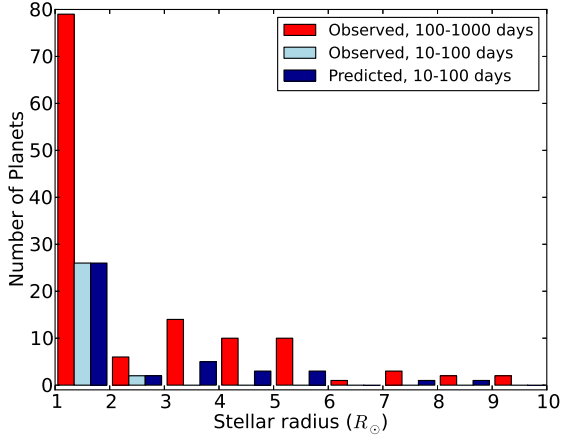}
\caption{\label{fig:WJnumbers}
 The red bars show the number of observed planets with orbital periods between 100--1000 days (LJ), as a function of host
star radius. The light blue bars show the equivalent WJ population (orbital periods 10--100 days). The dark blue bars represent
the expected WJ population if we simply scale the LJ population to match the overall occurrence rate. The lack of observed
(light blue) planets relative to expected (dark blue) planets is apparant for stellar radii $> 2 R_{\odot}$.
}
\end{center}
\end{figure}

\begin{figure}
\begin{center}
\includegraphics[width=1.0\columnwidth]{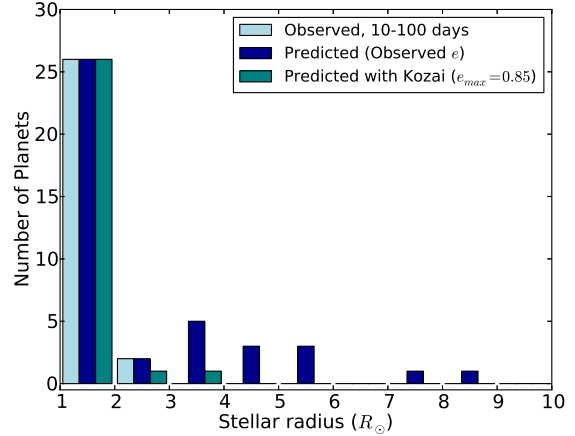}
\caption{\label{fig:WJnumbersKozai}
 We repeat the comparison of Figure~\ref{fig:WJnumbers}, comparing the observed (light blue) WJ population as a
function of host star radius to that expected from a scaled population of LJ (dark blue). We now also include a prediction
(green bars) which assumes that the scaled LJ population all oscillate in eccentricity on timescales short compared to the
stellar evolution timescale, with a maximum eccentricity = 0.85. We see that this removes most of the excess predicted
population at stellar radii above $2 R_{\odot}$.
}
\end{center}
\end{figure}

As illustrated by Figure \ref{fig:WJnumbers}, this calculation predicted a significant population (15) of observed WJs around evolved stars, which is inconsistent with the observed number (2). %What this discrepancy indicates is that if the eccentricity of WJs is as dominated by low values as 
If, however, each WJ is oscillating between some minimum and maximum value of eccentricity, then fewer will survive stellar expansion as they are removed at their minimum periapse (at maximum eccentricity), rather than the value currently observed.
 The actual maximum eccentricity will depend on a particular planet--perturber configuration, but we can assess the direction
of the effect by
 assuming all WJs are undergoing these oscillations up to a maximum eccentricity of $e_\text{max}=0.85$, from which calculated a predicted number (2) that equals the value from observations (Figure \ref{fig:WJnumbersKozai}).

\begin{figure}
\begin{center}
\includegraphics[width=1.0\columnwidth]{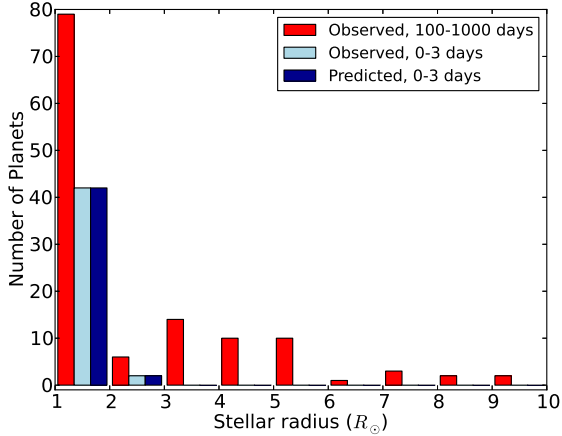}
\caption{\label{fig:HJnumbers}
The observed number of LJs and the observed and predicted number of very hot jupiters (Period$<3$ days) as a function of stellar radius. Unlike the predicted number WJs, the predicted number of these planets matches observations without any variation in the periapse distribution.%
}
\end{center}
\end{figure}

We make a similar estimate of the number of missing HJs using planets on periods $<3$ days, which are very unlikely to be oscillating given the strong tidal interactions at such short periods. Figure \ref{fig:HJnumbers} shows that the observed number of planets agrees with the number predicted by the current eccentricity distribution and does not benefit from a periapse distribution shifted to lower values. 

This brief calculation illustrates that the lack of WJs is unlikely to be a simple statistical fluctuation, and necessitates an explanation. We note that it does include a number of assumptions, foremost being that the relative frequency of WJs to LJs is independent of stellar radius outside of removal via tides. However, the purpose of this calculation is not to determine the precise number of WJs removed to due stellar evolution, but rather to demonstrate that an absence exists and can be accounted for they are oscillating to higher eccentricity values, as would be required for migration via tides. A more detailed analysis follows in Section \ref{sec:rstar}, with further discussion of assumptions in Section \ref{sec:observed}.
% Transit info
% Assuming all stars larger than 2 Rsun are evolved, and their orbiting exoplanets come from a population that is the same as for smaller stars.

\section{Dynamical effects}\label{sec:dynamics}

\subsection{The Kozai-Lidov mechanism}
The KL mechanism results from secular (long-term) interactions between an inner and outer body when their mutual inclination exceeds some nominal value, or if both bodies have significant eccentricity while coplanar \citep{Li:2013qy}. In the simplest case, where the inner body has negligible mass and the outer body is on a circular orbit, the z-component of the angular momentum is constant for the inner body:
\begin{align}
\cos i_\text{in}\sqrt{1-e_\text{in}^2} = \text{Const}
\end{align}
where $i_\text{in}$ is the inclination of the inner body (identical to the mutual inclination in the massless case) and $e_\text{in}$ is its eccentricity \citep{Kozai:1962qy, Lidov:1962uq}. 

This relationship requires that a decrease in mutual inclination between the two bodies is accompanied by an increase in the eccentricity of the inner body. As a result, the inner body undergoes oscillations in eccentricity and inclination under the influence of the outer companion. In this simple case these oscillations are characterised by their time-scale ($P_\text{Kozai}$) and maximum eccentricity \citep{Lidov:1962uq, Kiseleva:1998fk}:

\begin{align}
P_\text{Kozai} &= \frac{2}{3\pi}\frac{P_\text{out}^{2}}{P_\text{in}}\frac{M_\text{tot}}{M_3}(1-e_\text{out}^2)^{3/2}\\
e_\text{calc} &=\sqrt{1 - (5/3)\cos^2i_0}
\end{align}
where $P_\text{in}$ and $P_\text{out}$ are the inner and outer periods, respectively; $M_3$ and $M_\text{tot}$ are perturber mass and total mass of all the bodies, respectively; $e_\text{out}$ is the perturber eccentricity; $i_0$ is the minimum value of $i_\text{in}$; and $e_\text{calc}$ is the calculated maximum eccentricity. The maximum eccentricity has a more complicated, non-linear form when the inner body is massive.  \citet{Naoz:2013vn} derive the equation in the case of no initial eccentricity and a perturber on a circular orbit: 
\begin{align}\label{eq:emax}
\left( \frac{L_1}{L_2}\right)^4 e_\text{calc}^2 + \left( 3 + 4\frac{L_1}{L_2}\cos i_0 +\left(\frac{L_1}{2L_2} \right)^2\right)e_\text{calc}^2\nonumber\\
+\frac{L_1}{L_2}\cos i_0 - 3 + 5 \cos^2i_0=0
\end{align}
where $L_1$ and $L_2$ are the scaled angular momenta of the inner and outer orbit, respectively. These are defined as: 
\begin{align}
L_1 &=\frac{M_1M_2}{M_1+M_2}\sqrt{G(M_1+M_2)a_\text{in}} \\
L_2 &=\frac{M_3(M_1+M_2)}{M_1+M_2+M_3}\sqrt{G(M_1+M_2+M_3)a_\text{out}}
\end{align}
where $M_1$, $M_2$, and $M_3$ are the masses of the central body, inner body, and perturber, respectively, and $a_\text{in}$ and $a_\text{out}$ are the inner and outer semi-major axes (SMA). Including the initial eccentricities of both orbits modifies the calculation only slightly and gives a value that differs at most by a few percent.

%white dwarf mergers Prodan:2013lr

The KL mechanism has been covered extensively in the literature in a number of contexts, including asteroids \citep{Fang:2012ys}, exoplanet systems \citep{Naoz:2011yq, Petrovich:2015qe}, the dynamics of the Galactic Center \citep{Lockmann:2008vn}, and stellar triple systems \citep{Eggleton:2001lr, Fabrycky:2007dz, Thompson:2011th, Prodan:2013lr, Naoz:2014bs}. Recently it has been shown that the inclusion of higher-order terms can dramatically alter the oscillations induced by the KL mechanism. These octupole terms, which are non-zero if the inner body is not massless or the outer body has a non-zero eccentricity, can result in larger eccentricities, flips of the inner orbit, and chaotic behavior \citep{Katz:2011dq, Lithwick:2011bh, Naoz:2013vn}. The full equations have no analytical solution for the maximum eccentricity of the inner orbit. However, Equation \ref{eq:emax} still provides an adequate first-order estimate of $e_\text{max}$, which we use in Section \ref{sec:setup} to limit our parameter space to those systems that could be capable of migrating inward.  

In the absence of any other effects, oscillating bodies can approach within an arbitrary distance of the surface of the star as long as they avoid collision. However, two effects prevent that from happening: general relativity (GR) and tides. 

\subsection{General relativistic precession}\label{sec:GR}
For short-period orbits, GR causes a precession of apsides on a time-scale that depends on the properties of the orbit and the host star:

\begin{align}
P_\text{GR} = \frac{P_\text{in}^{5/3} c^2(1-e_\text{in}^2)}{3(2\pi^{5/3})(GM_1)^{2/3}}
\end{align}

If $P_\text{Kozai}$ is longer than this time-scale, GR precession can damp and eliminate KL oscillations. For this reason, orbits with shorter periods require stronger perturbers to undergo oscillations: those that are more massive, closer, and/or more eccentric. As shown in \citet{Dong:2014zr}, WJs need perturbers within 10 au to undergo the high eccentricity migration discussed  here. This constraint is due in part to the eccentricity dependence of the GR time-scale. Without a strong enough perturber, oscillating bodies that reach very high eccentricities can be stranded at their maximum eccentricity. When tides are taken into account, that can lead to rapid evolution into a HJ. While the detailed effects of GR are significantly more complicated \citep{Naoz:2013fj}, the damping interpretation is adequate for our purposes.

\subsection{Tides}
\subsubsection{Tidal decay}\label{sec:tidaldecay}
Both the KL mechanism and GR conserve orbital energy, ensuring that the SMA of the inner orbit is constant. WJs can only migrate when under the influence of a dissipative force, which takes the form of tidal friction. The existence of WJs at their current periods, as well as their non-zero eccentricities, indicate that they must have large circularization and tidal decay time-scales as a result of weak tidal forces. In this section we describe tidal forces that hold for two bodies in general, but in our simulations apply specifically to the planet and star.

The effects of tidal forces on orbital evolution (first investigated in the context of planets and satellites, see \citealt{Darwin:1880qf}) have  been investigated in detail for stars, showing that tides reduce orbital energy and lead to smaller and more circular orbits \citep{Hut:1981pd, Eggleton:1998ve, Kiseleva:1998fk}. For two tidally interacting bodies, whether massive planets or stars, the strength of tides raised on an object 1 by an object 2 is characterised by the tidal friction time-scale, as described in \citet{Eggleton:2001lr} and \citet{Fabrycky:2007dz}:
\begin{align}\label{eq:tfriction}
t_{F1} = \frac{t_{V1}}{9}\frac{a^8}{R_1^8}\frac{M_1^2}{(M_1 + M_2)M_2}(1+2k_1)^{-2}
\end{align}
where $a$ is the SMA of the orbit and $R_1$ is the radius of object 1. The internal structure of object 1 is included by way of $k$, the classical apsidal motion constant, which represents the quadrupolar deformability of the star or planet, and $t_V$, the viscous time-scale, which is a parametrization of internal dissipation in the star \citep{Zahn:1977lr}. The physical values parametrizing tidal evolution are still not fully understood, although they have been investigated by a number of authors. The planetary $k$ is frequently set to $k_\text{P}=0.25$, the result for a $n=1$ polytrope representing gas giants. Recent research has gone into matching $t_V$ to observations, including the Jupiter-Io system, the eccentricity distribution of hot Jupiters, and the existence of high eccentricity exoplanets. \citet{Hansen:2010lq} calibrated tidal models to observations of massive exoplanets (0.3-3 $M_J$) around solar-type stars using a single tidal dissipation constant for each population, and found that $t_V$ for a Jupiter-mass planet with moderate eccentricity is $t_{Vp}=150$ years.
 However, longer orbital periods may imply greater dissipation as they couple to a larger fraction of the internal turbulent
viscosity. \citet{Hansen:2012ly} finds a roughly linear increase in the dissipation rate with orbital period. Furthermore,
 \citet{Socrates:2012mz} repeated a similar calibration in the high eccentricity limit and determined that planets undergoing high-eccentricity migration require  tides equivalent to $t_{Vp}=1$ year. It is this value we use in the numerical simulations of Section \ref{sec:results}. 

\citet{Hansen:2010lq} also found that tides raised on the solar-type host stars by Jovian-mass planets were a factor of 50 weaker than those raised on the planets by the stars, allowing us to ignore stellar tides for our numerical simulations. However, this inequality does not hold as stars evolve. The strong radius dependence of $t_{F*}$ indicates that as a star leaves the main sequence it will increase its contribution to the planet's orbital evolution until stellar tides dominate or the star engulfs the planet. Once the star dominates tidal effects the strong radius dependence will rapidly accelerate the inward migration of the planet. Whether this increase in migration rate occurs before direct collision with the star depends on the migration rate due to planetary tides alone, as discussed in greater detail in Section \ref{sec:planetremoval}. When stellar tides are included, we use the values $k_*=0.014$, based on an $n=3$ polytrope, and $t_{V*}=50$ years, from the equation provided in \citet{Eggleton:2001lr}. Stellar tides are likely weaker than this value, as seen by the results of \citet{Hansen:2010lq}, but our choice of $t_{V*}$ does not affect our conclusion as long as it is longer than $t_{Vp}$.

\subsubsection{Rotational effects}\label{sec:tidesrot}
Tidal forces also exert a torque on a planet, changing its spin and aligning it on time-scales much shorter than those required to circularize the orbit or move the planet inward. Planetary systems residing in the WJ period range as a result of migration should have reached an equilibrium in their spin as a result of this effect. In the case of planets not undergoing oscillations in eccentricity, the equilibrium spin can be determined by the value which results in no torque, or pseudo-synchronous (PS) spin \citep{Hut:1981pd}. In the case of low eccentricity, the planetary spin period is the same as its orbital period (synchronous rotation). For large values of eccentricity, the planet is moving much more rapidly at periapse, where tidal forces are strongest, and as a result the planet rotation period can be less than 1 percent of the orbital period. Those planets rotating faster than the PS value will have angular momentum transferred from its rotation to its orbit, which can result in a modest increase in SMA.

\subsection{Stability}
Finally, for these three-body systems to exist they must be stable. While KL oscillations require a strong perturber to avoid damping by GR, a perturber that is too near to the inner orbit will destabilize the system. The limit for stability in mutually inclined systems with an eccentric perturber was calculated by \citet{Mardling:2001cr}:
\begin{align}\label{eq:stability}
\frac{a_\text{out}}{a_\text{in}}>2.8(1+q)^{2/5}\frac{(1+e_\text{out})^{2/5}}{(1-e_\text{out})^{6/5}}\left(1-0.3\frac{i_\text{tot}}{180^\circ}\right)
\end{align}
where $q=M_3/(M_* + M_p)$. This criterion has been used in prior investigations of KL oscillations in exoplanet systems, including \citet{Teyssandier:2013fj} and \citet{Rice:2015rt}.

With these effects in mind, the planetary systems we want to investigate are those that are undergoing KL oscillations, requiring $P_\text{Kozai} <P_\text{GR}$ over the full range of eccentricities that the planet reaches. The maximum eccentricity due to oscillations should be large enough to induce tidal decay, but over a time-scale large enough that a population of WJs would be detectable. 

%\subsection{The relevant time-scales}
%
%In order to specify whether a given system has a planet migrating inward, we need to define the five important time-scales:
%\begin{itemize}
%\item $P_\text{in}$: Inner orbital period
%\item $P_\text{out}$: Outer orbital period
%\item $P_\text{KL}(e_\text{out})$: Kozai-Lidov time-scale
%\item $P_\text{GR}(e_\text{in})$: GR precession time-scale
%\item $P_\text{tides}(e_\text{in})$: Tidal time-scale
%\end{itemize}
%
%Few observed warm jupiters have known companions, meaning the outer orbital period and the KL time-scale unknown in those cases.
%
%For a WJ to be in the process of migrating inward due to the KL mechanism, two relationships need to hold:
%\begin{eqnarray}
%P_\text{KL} &< P_\text{GR} \\
%P_\text{tides} &> *what*
%\end{eqnarray}
%
%
%

\section{Numerical simulations}\label{sec:setup}
% Why are we running these simulations: to understand how the mass, eccentricity, and period affect the da/dt in the case of Kozai-Lidov oscillations in WJs. Also to see what the maximum eccentricities reached are, compared to the calculated values, and to get the eccentricity distribution as a function of time. 
The orbital evolution of an oscillating planet depends on both the maximum eccentricity and the distribution of eccentricity values over time. These properties of the system do not have an analytical form. The maximum eccentricity deviates from $e_\text{calc}$ in Equation \ref{eq:emax} due to octupole terms, while the eccentricity distribution has no analytic form even without octupole terms. In order to understand the orbital evolution of migrating WJs, we need to use numerical simulations spanning the parameter space of interesting systems. These simulations allow us to determine the orbital decay as a function of initial period, mass, and  eccentricity, as well as determine the relationship between calculated ($e_\text{calc}$) and true maximum eccentricity.  In our simulations we use
 the code of S. Naoz, which integrates the three-body secular equations up to the octupole level of approximation
 as described in \citet{Naoz:2013vn}, including GR effects for the inner and outer orbits and tidal effects following \citet{Eggleton:2001lr} and \citet{Fabrycky:2007dz}. This code has been used extensively in numerous calculations e.g. \citet{Naoz:2011yq, Naoz:2012pd, Naoz:2014bs, Li:2014rr, Li:2015fy}.

\subsection{Creating a population}
Each system is composed of a central star, an inner body (hereafter referred to as ``planet"), and an outer body (hereafter referred to as ``perturber"). For our star, we selected a mass of 1.2 $M_\odot$ and ignored the contribution of tides raised on the star to the  evolution of our planetary orbit (see Section \ref{sec:tidaldecay}). The other properties of our systems were chosen to produce all three of the following properties in the planet: 
\begin{itemize}
\itemsep.3em
\item Warm ($P=10-100$ days) Jupiters ($M_p = 0.1-10 M_J$) 
\item Undergoing KL oscillations
\item Experiencing  tidal migration on a plausible timescale
\end{itemize}
Each of these  requirements introduces constraints onto the population.  The first constrains the mass and period of the planets,  the second constrains the perturber such that  GR time-scale is longer than the KL time-scale, and  the third constrains the planet to reach high eccentricities during oscillations. The consequences for perturber and planet properties are discussed below.

\subsubsection{Perturber properties}\label{sec:perturberprops}
% While perturber does play a pivotal role in the modulation of the planetary orbit, 
We limited the parameter space of our primary simulations by keeping the perturber constant across them. We selected its properties such that it caused KL oscillations in the systems  with semi-major axis $\sim0.1$ au, which were most sensitive to quenching by GR (Section \ref{sec:GR}, constraint 2 above), while avoiding system instability in the largest orbits ( semi-major axis of 0.45 au). We arrived at a 4 $M_J$ body at 2 au with an eccentricity of 0.13, similar to the planets that are sometimes detected as companions to HJs \citep{Knutson:2013mz}. Plugging these numbers into Equation \ref{eq:stability}, we find that the limit for stability is $a_\text{out}/a_\text{in}>3.5$. Our systems have $a_\text{out}/a_\text{in} =  4.4-20$, clearly in the stable regime. Additionally, we ran two other sets of simulations: one with simply a larger eccentricity (0.35), and one with a larger ($30 M_J$), more eccentric ($e=0.64$) perturber at a larger distance (10 au), both discussed in Section \ref{sec:varyperturber}. These simulations showed that while the perturber plays a pivotal role in the planetary oscillations and migration, the perturber properties did not impact our general results.

%We note that recently \citet{Ngo:2015ys} showed that a stellar companion between 50 and 2000 au does not change the likelihood of HJs being eccentric or misaligned, indicating that if the KL mechanism is accountable for a significant fraction of HJs then the perturber needs to be a close stellar companion or another planet. The perturbers in our simulations are located at distances, 2 and 10 au, that are consistent with that result.

\subsubsection{Inner planet properties}
We defined our population of WJs to  have masses $0.1- 10 M_J$ and  orbital periods from $10 - 100$ days, or  semi-major axes $0.1-0.45$ au. For simplicity, we set the size of all planets to 1 Jupiter radius, with $k_p=0.25$ and $t_V=1$ year as described in Section \ref{sec:tidaldecay}. While planets at the low-mass end of our population are unlikely to be this large, there is not a firm mass-radius relationship for extrasolar planets at this point. We discuss the impact of this assumption in Section \ref{sec:masseffect}. To generate the properties of our planet, we first randomly sampled the mass range, initial eccentricity, and mutual inclination. We did so logarithmically in mass and uniformly in  initial eccentricity and inclination, limiting the latter to $0-0.1$ and the former to $70^\circ-90^\circ$. We chose these boundaries based on the properties of the Kozai-Lidov oscillations which are needed to match our requirement that the minimum eccentricities be close to circular -- to match observations -- while still allowing planets to reach sufficiently large eccentricities to evolve tidally on a short enough timescale.

\begin{figure}
\begin{center}
\includegraphics[width=0.98\columnwidth]{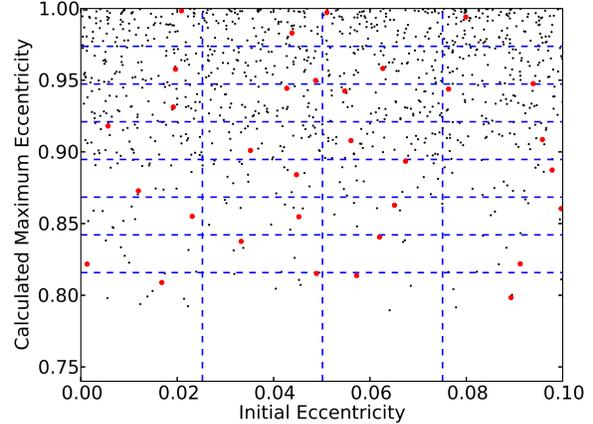}
\caption{\label{fig:MCemaxemin}
A randomly generated distribution of planets based on our limits (black points), along with the bins (blue lines), and a set of selected systems (red dots). This approach ensured we simulated a wide range of systems rather than be dominated by the portions of parameter space with the majority of points.}
\end{center}
\end{figure}

From these properties and those of the perturber, we calculated $e_\text{calc}$ for all samples using Equation \ref{eq:emax}. We then selected a uniform distribution in initial eccentricity and $e_\text{calc}$ by dividing the parameter space up into a grid and selecting systems from each grid box, as shown in Figure \ref{fig:MCemaxemin}. This approach allowed us to probe the wide range of behavior caused by different minimum periapse values while still limiting computation time. We constrained the initial eccentricity to between 0 and 0.1 and $e_\text{calc}$ between 0.75 and 1.0 to produce oscillating systems that had eccentricity values enabling migration. While $e_\text{calc}$ is only accurate for systems without any contribution from octupole terms, it gave us a first approximation and allowed us to exclude systems that are unlikely to migrate. We also set both arguments of periapse to zero in order to limit our parameter space, as other groups have done \citep{Teyssandier:2013fj}. Generally speaking, this assumption is equivalent to maximizing the effect of the companion, leading to the largest peak eccentricity. 

\subsubsection{Planetary rotation}\label{sec:prot}
As described in Section \ref{sec:tidesrot}, planetary rotation can have a significant effect on orbital evolution via tides.  We tested this by running a set of simulations with a range of planetary rotation rates and found that as expected planets spinning faster than the equilibrium (PS) rate migrated more slowly, or in some cases migrated outwards. Because these systems are assumed to be migrating WJs, which originated beyond periods of 100 days, they should have already reached PS rotation. For our main simulations, we used $e_\text{calc}$ to estimate the PS rotation period. However, the  planets nearest to the star (10 and 20 days) reached maximum eccentricity values significantly different from $e_\text{calc}$, due to GR and tidal effects, resulting in planets  spinning too rapidly. To correct for this, we performed a linear fit between the calculated and simulated maximum eccentricity, and used the derived eccentricity value to set the correct rotation rate. We also set a lower limit on the spin period by capping the eccentricity used in its calculation to $1 - 2R_\odot/a$, the value that brings the planetary periapse to $2R_\odot$. This limit avoided spin rates that were unreasonably fast, exceeding the maximum physical rotation rate of a Jupiter-mass planet.

\subsection{The full population}
We ran a total of 1,320 simulations across 6 period values: 10, 20, 30, 50, 70 and 100 days. We also ran another 384 with different perturbers and 192 with reduced viscous time-scale, both across the same period range.
 The goal is not to simulate the transition from WJ to HJ, because we are interested in those which have not yet completed such a transition on an astrophysical timescale yet are evolving fast enough to have moved to their current locations.
% The number of simulations was chosen based on computational constraints, but effectively probed the parameter space (see Figure \ref{fig:peridadt}). 
All simulations  were run for $10^6$ years or until the orbit of the planet decayed by 10 percent, whichever occurred first. While this is short compared to the full migration time it is long enough to encapsulate many eccentricity oscillations and thus characterise the rate of migration at the observed stage.

For comparison to systems not undergoing oscillations, we also ran an additional set of 192 simulations over the same period bins without the effects of a perturber. These simulations spanned both the same mass range (3 bins: 0.1, 1, 10 $M_J$) and the full eccentricity range (19 bins from 0 to 0.95), and began with the theoretical value for PS rotation at their eccentricity. 

\section{The results}\label{sec:results}

%%%%% figure out how to organize this
\subsection{A single system}
\begin{figure}
\begin{center}
\includegraphics[width=1.0\columnwidth]{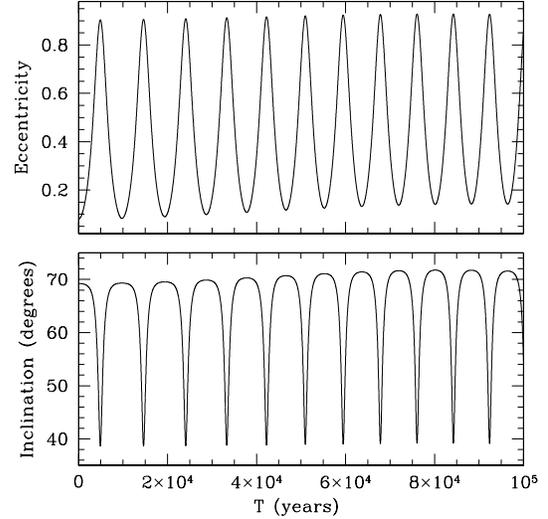}
\caption{ The upper panel shows the eccentricity evolution and the lower panel shows the inclination evolution. This
shows the characteristic sweep from the initial low eccentricity to very high values, as the inclination falls to its
minimum.
\label{fig:snapshot}}
\end{center}
\end{figure}

As a case study, we selected a system with $M_p = 0.6 M_J$ at 50 days (0.28 au), with an initial eccentricity of $e=0.08$ and an initial mutual inclination of $i_0=72.2^\circ$. Using Equation \ref{eq:emax} we determined $e_{calc} = 0.91$, which is large enough to  drive inward migration as long as the planet is not spinning extremely rapidly. This eccentricity corresponds to a periastron of $5R_\odot$ at closest approach, well outside of the $2R_\odot$ limit we set for PS rotation calculations in Section \ref{sec:prot}. We set the rotation period to the calculated value of $P_\text{rot}/P_\text{orb} = 42$, or $P_\text{rot}=1.2$ days. 
%As a case study, we pick a system with $M_p = 2.2 M_J$ at 50 days (0.28 au), with an initial eccentricity of $e=0.04$ and an initial mutual inclination of $i_0=86.8^\circ$. Using Equation \ref{eq:emax} (OR VERSION FROM SMADAR'S PAPER) we calculate the predicted maximum eccentricity to be $e_{calc} = 0.98$, large enough that we expect some inward migration as long as the planet is not spinning extremely rapidly. However, this eccentricity would bring the planet to within $\sim15$ percent of the stellar radius, which is unlikely due to the effects of tides and GR. To calculate the planetary spin we instead use $e=0.965$ as described in \ref{sec:prot} to get $P_\text{rot}/P_\text{orb} = 5.6\times10^{-3}$, or $P_\text{rot}=6.7$ hours.
 Figure~\ref{fig:snapshot} shows the resulting eccentricity and inclination evolution of the inner planet, with a Kozai oscillation period of $3.2 \times 10^3$~yrs, from a direct numerical simulation.

\begin{figure}
\begin{center}
\includegraphics[width=1.0\columnwidth]{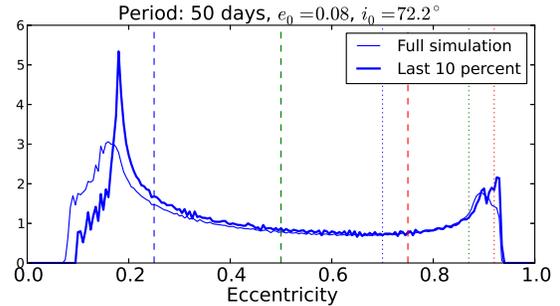}
\caption{Eccentricity distribution of our 50-day case study planet, migrating inwards with $\Delta a/\Delta t=-3$ au/Gyr, during the full $10^6$ year simulation (thin line) and the final $10^5$ years (thick line). The vertical lines show the 75th, 90th and 98th percentile of the distribution of observed WJ eccentricities (dashed, from the Exoplanet Orbit Database) and our simulated eccentricity distribution (dotted). We see that, for this case, the simulated eccentricity distribution is biased high relative to the observed one.\label{fig:caseecc}}
\end{center}
\end{figure}

 Figure~\ref{fig:caseecc} shows the relative fraction of the time spent at each eccentricity when the simulation is run for $10^6$ years.
We found that the eccentricity peaked at a value of 0.93, slightly larger than our calculation (Figure \ref{fig:caseecc}). Additionally, the difference between the full $10^6$ year distribution and the distribution in the last $10^5$ years (thick line) showed that the minimum eccentricity during oscillations increased slightly over time. Given the distribution of eccentricity as a function of time, a system with these properties would most likely be observed with $e<0.4$, but would be detected 25 percent of the time with $e>0.7$. Migrating inward at 3 au per Gyr, such a planet would survive for significantly less than the migration time-scale $a/\dot{a}=0.094$ Gyr, due the increase in the strength of tides as the SMA gets smaller. A migration time-scale of this duration is short compared to the ages of WJ host stars, indicating that this type of system could have migrated to its current location from farther out. 

% 50 day alt, v11, e0_0.08_i0_1.26_emax_0.91

\subsection{All systems}
\begin{figure*}
\centering
\hspace{0.32in}
\includegraphics[width=5in]{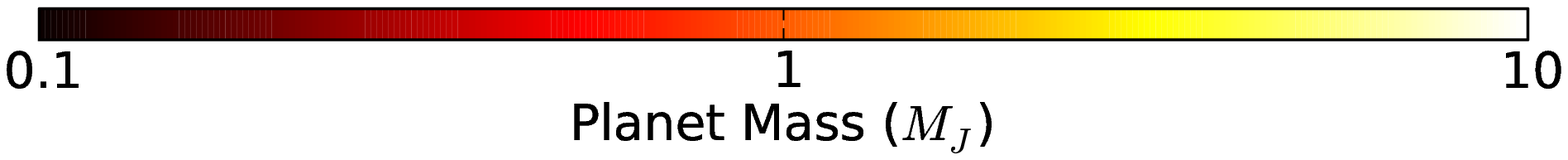}\\
\vspace{-0.12in}
\begin{tabular}{cc}
	\vspace{-.22in}
%	\vspace{-.07in}
	\hspace{-.4in}
	\includegraphics[width=3.8in]{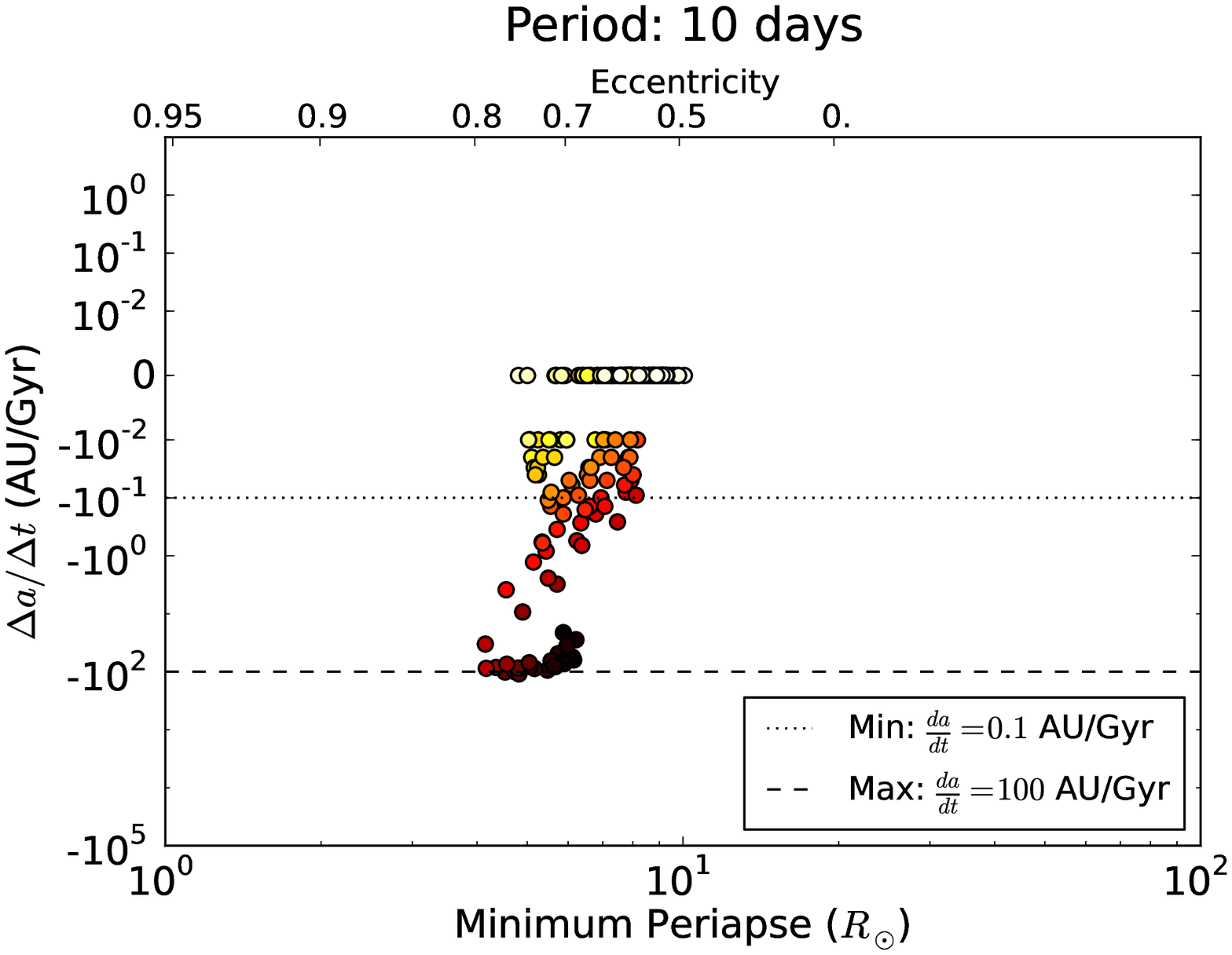} &
	\hspace{-.3in}
	\includegraphics[width=3.8in]{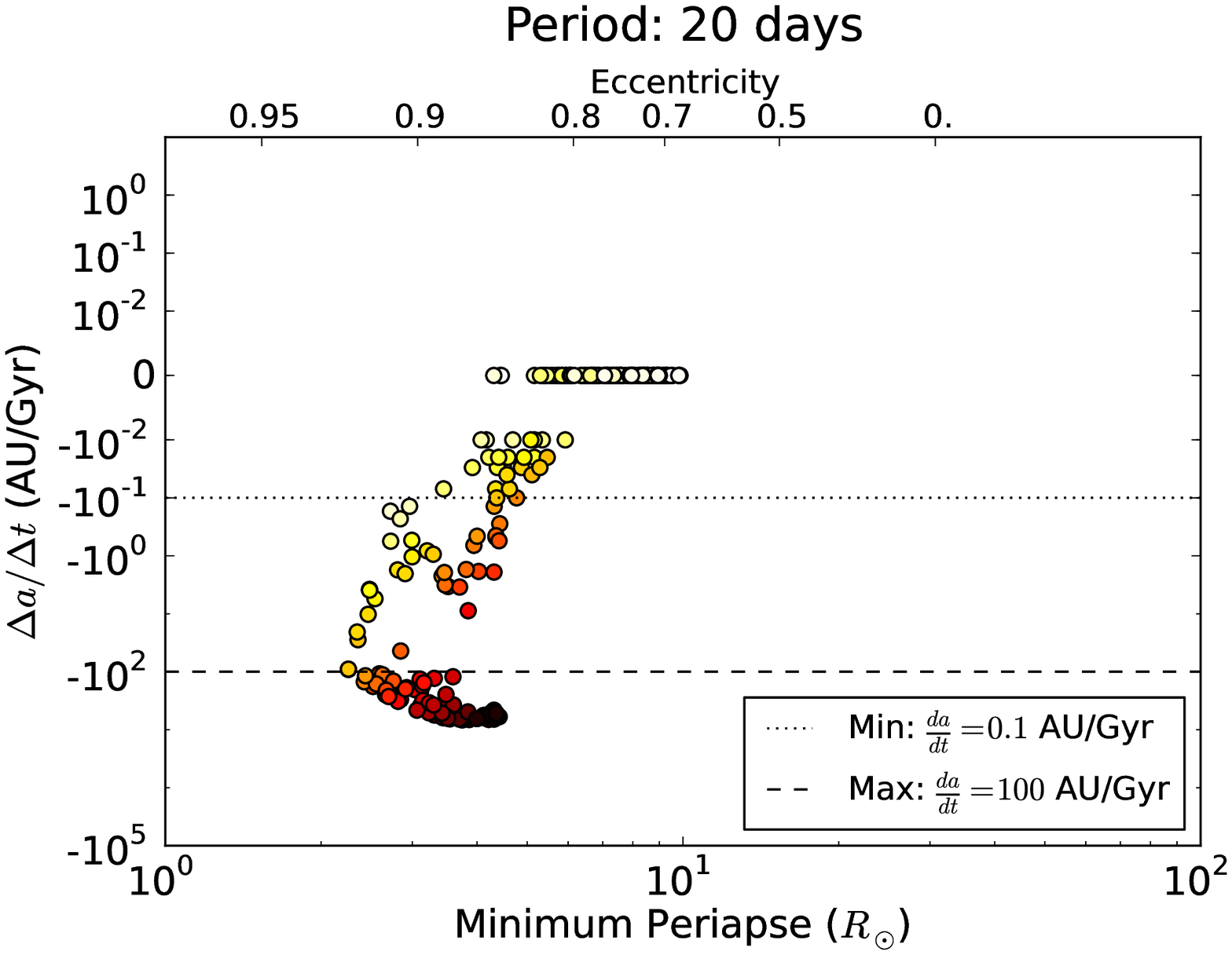} \\
%	(a) & (b)\\[6pt]
%	\vspace{-.07in}
	\vspace{-.22in}
	\hspace{-.4in}
	\includegraphics[width=3.8in]{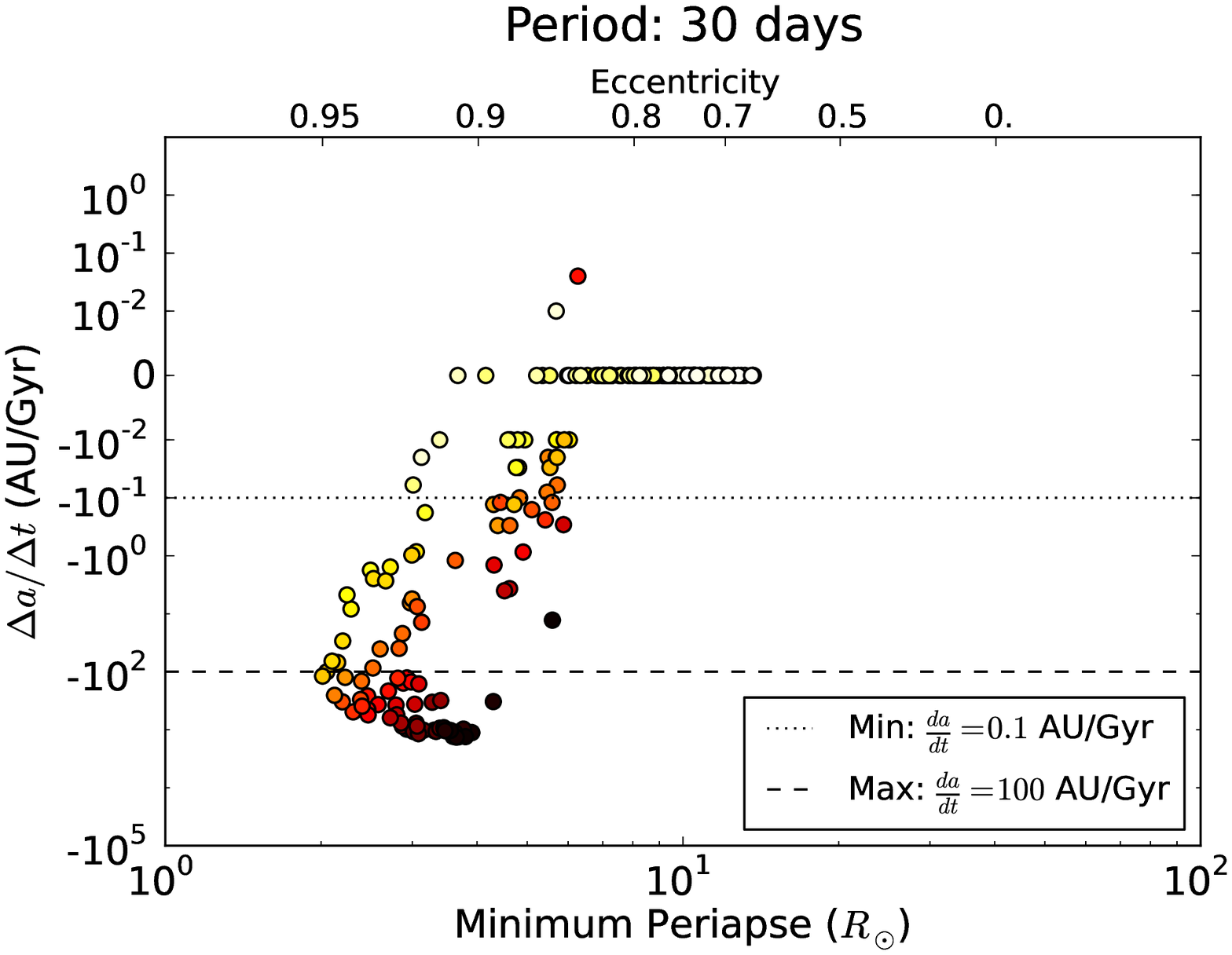} &
	\hspace{-.3in}
	\includegraphics[width=3.8in]{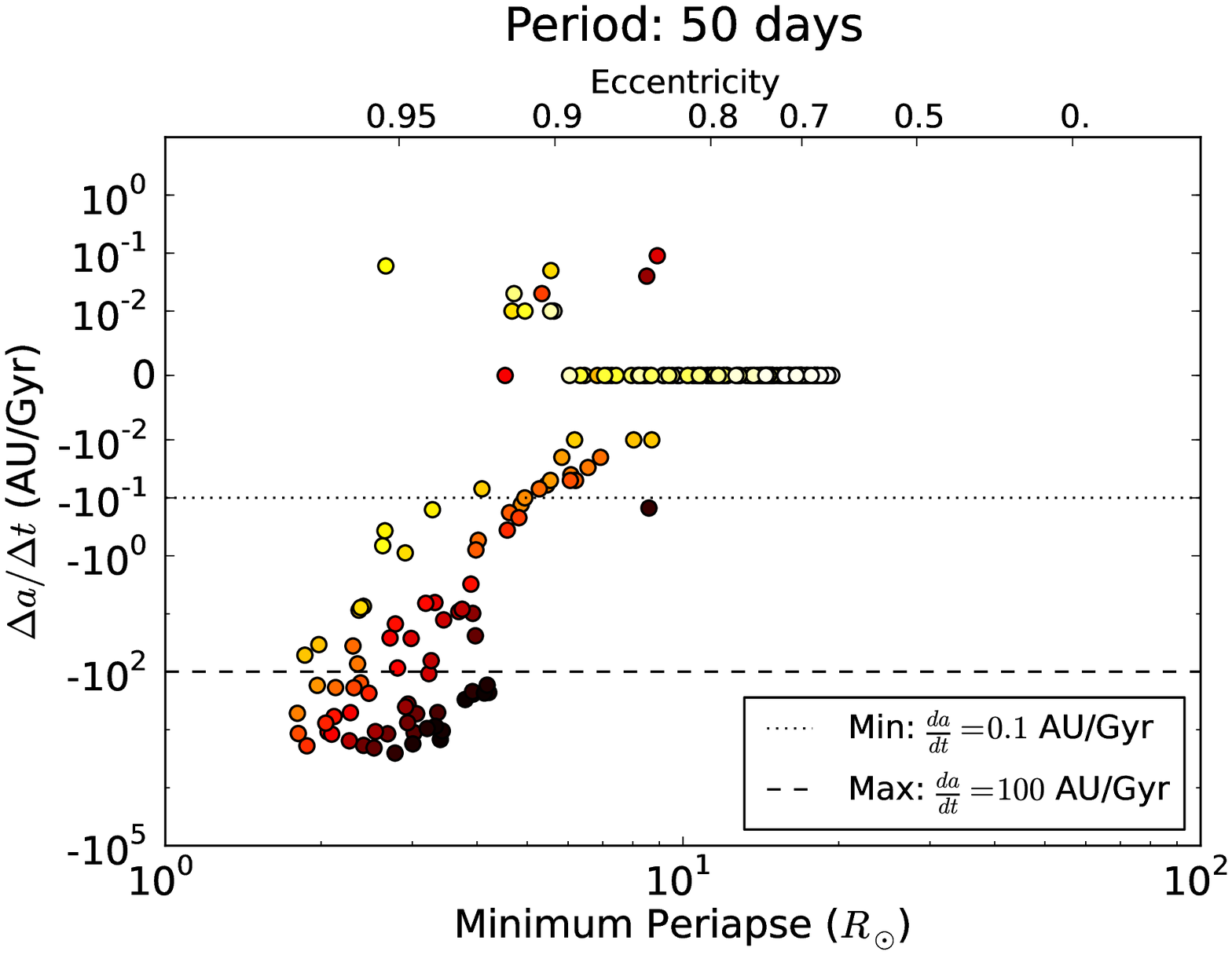} \\
%	(c) & (d)\\[6pt]
	\vspace{-.07in}
	\hspace{-.4in}
	\includegraphics[width=3.8in]{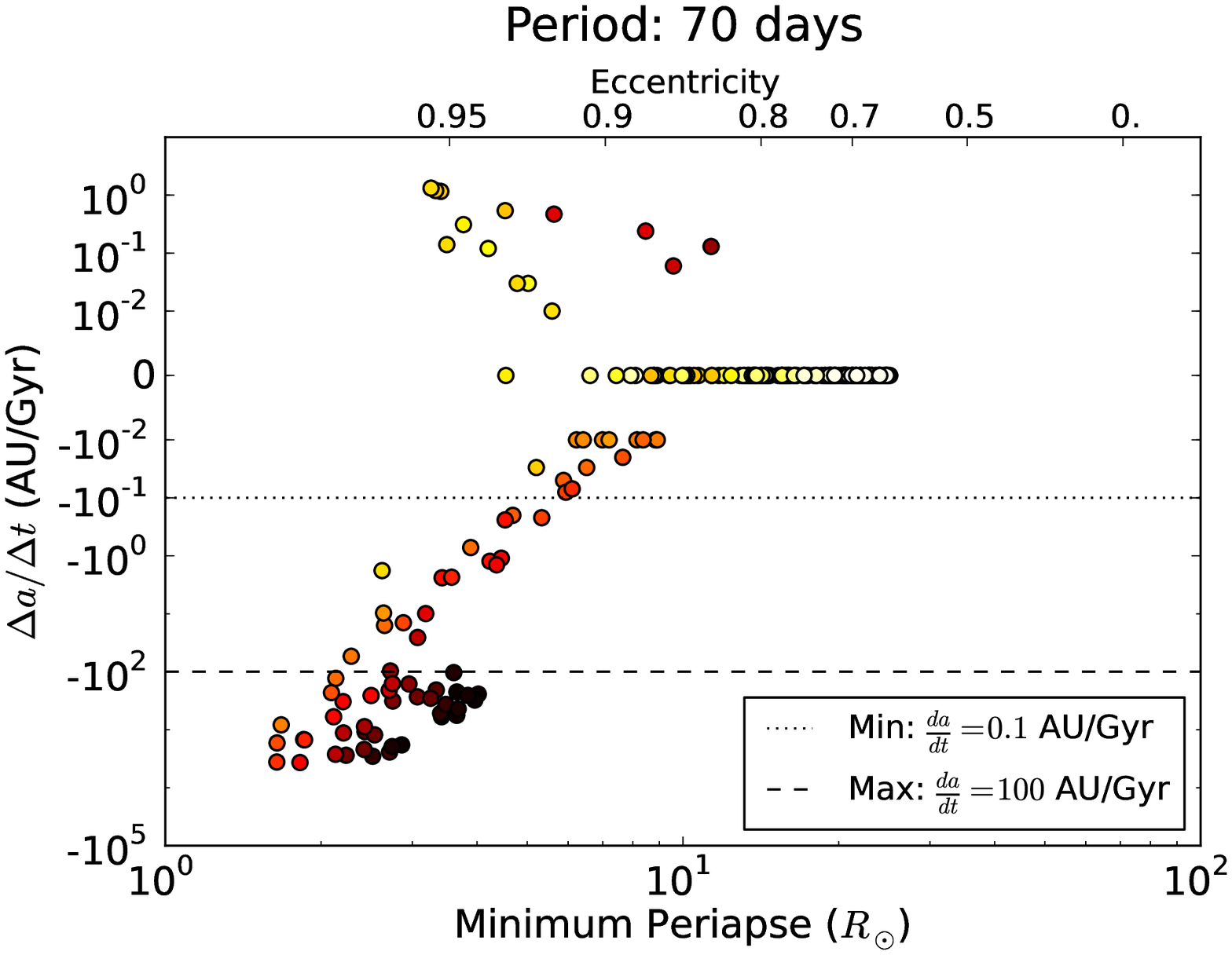} & 
	\hspace{-.3in}
	\includegraphics[width=3.8in]{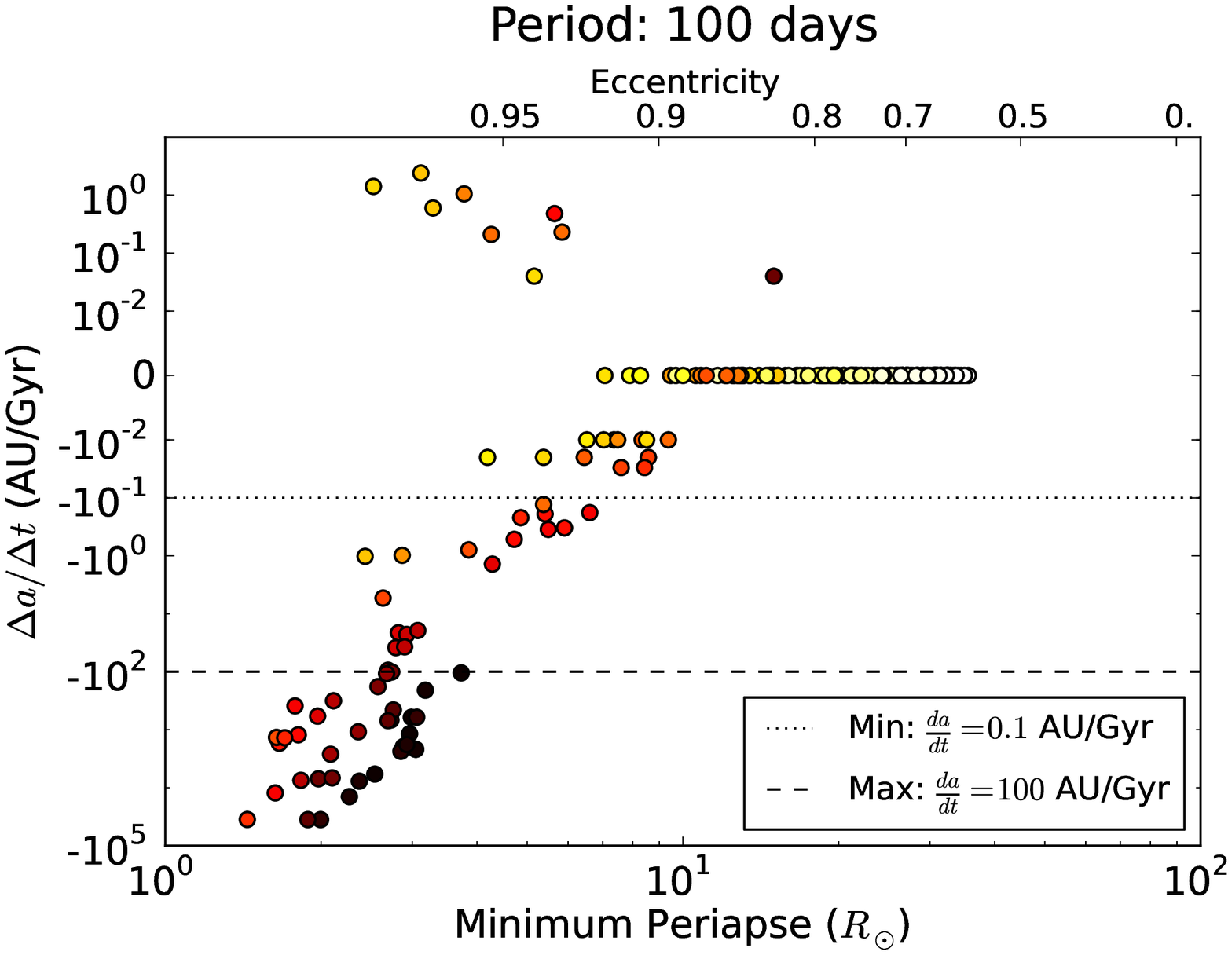}\\
%	(e) & (f)\\[6pt]
\end{tabular}
\caption{The change in semi-major axis over the duration of  each simulation,  grouped by orbital period. The x-axis is minimum periapse/maximum eccentricity while the color of the points gives the mass of planet.
 The dotted (maximum) and dashed (minimum) black lines correspond to the approximate limits on migration rate necessary to produce  the WJ population. If they migrate too fast, they will be observed as HJ, and if they migrate too slowly, they will remain in the LJ population.
The blue dotted line indicates the tidal disruption radius. \label{fig:peridadt}}
\end{figure*}

\begin{figure}
\begin{center}\includegraphics[width=1.0\columnwidth]{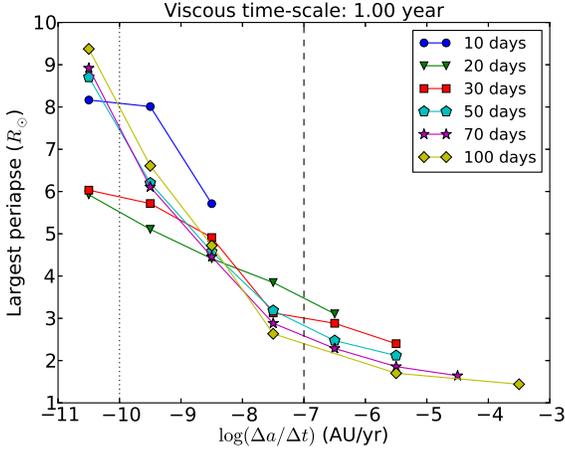}\caption{The largest minimum periapse that resulted in the planet migrating at a given rate, for each orbital period. The systems used in Section \ref{sec:rstar} fall between the dotted and dashed lines.  \label{fig:largestperi}}
\end{center}
\end{figure}

Repeating this process on all 1,320 systems produces the results shown in Figure \ref{fig:peridadt}: an instantaneous migration rate ($\Delta a/\Delta t$) as a function of minimum periapse/maximum eccentricity and planetary mass. These plots illustrate the extremely strong dependence of migration rate  on maximum eccentricity, as expected. In all simulations with planet periods longer than 10 days, orbital migration only took place when the maximum eccentricity exceeded 0.8. This result is significant, as all observed WJs have eccentricities below this value (see Figure \ref{fig:obsWJeccdist}), which will be discussed in Section \ref{sec:observed}.  Notably, the same periapse distance results in similar migration rates regardless of period (Figure \ref{fig:largestperi}). This result plays an important role when determining how the population of WJs is affected by stellar evolution in Section \ref{sec:rstar}. The partial exception to this phenomenon are those planets at 10 day periods, which are near enough to their host to experience tidal effects with even moderate eccentricity. 

% On each plot we indicate two migration rates, which bound our ``interesting'' sample. These are systems that are migrating between 0.1 and 100 au per Gyr. Planets migrating more rapidly than this are unlikely to represent any observed population, unless there is a very large number of such planets migrating, establishing a significant steady-state flow. However, as 

Additionally, some systems at larger periods migrated outward rather than inward. In these cases our estimate for the rotation rate was too high, possibly due to the effect of octupole terms in the KL oscillations or tidal effects, and they experienced outward migration due to their spin down. The relative symmetry between the outward and inward moving planets is due to the magnitude of migration being set primarily by the product of the tidal friction time-scale (Equation \ref{eq:tfriction}) and a function of eccentricity dominated by a $(1-e^2)^{-13/2}$ coefficient. We also note that none of our systems spent any time on retrograde orbits. This result is in line with the findings of \citet{Teyssandier:2013fj}, which showed that highly inclined systems are significantly poorer at causing flips in the planet. In our simulations only highly inclined systems produce large eccentricities, and as a result all stayed prograde. 

\subsubsection{Eccentricity frequency distributions}
\begin{figure}
\begin{center}
\includegraphics[width=1.0\columnwidth]{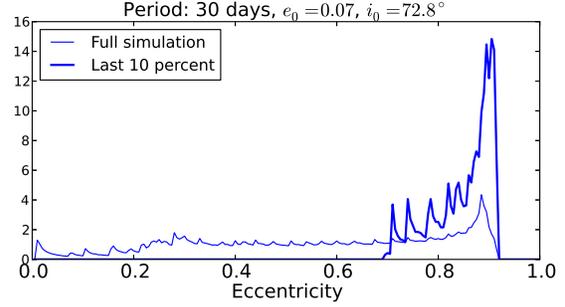}
\caption{Eccentricity distribution of a rapidly migrating planet at 30 days ($da/dt=-4\times 10^{2}$ au/Gyr), illustrating that the majority of time is spent at high eccentricities during the last 10 percent of simulation time. This planet migrates too fast to be a plausible WJ candidate as it should rapidly circularise into a HJ orbit.\label{fig:bigecchist}
}
\end{center}
\end{figure}

\begin{figure}
\begin{center}
\includegraphics[width=1.0\columnwidth]{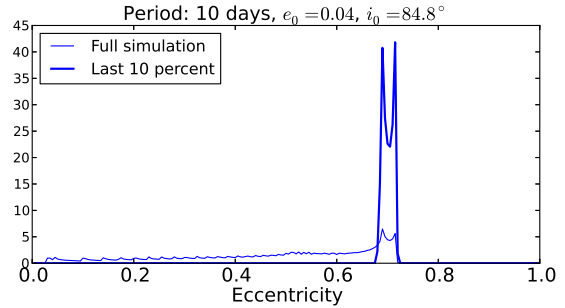}
\caption{Eccentricity distribution of a planet at 10 days with damped eccentricity oscillations, where the KL time-scale is roughly equal to the GR time-scale.  \label{fig:smallosc}}
\end{center}
\end{figure}

The eccentricity distributions of individual planets, along with the setup of the system, determined the magnitude of migration. Systems rapidly migrating ($da/dt>10^{2}$ au/Gyr) tended to peak more strongly at the high-eccentricity value, as seen in Figure \ref{fig:bigecchist}. Systems migrating on smaller time-scales ($10^{-1}-10^{2}$ au/Gyr) generally peaked near $e=0-0.2$ with a smaller additional peak between 0.8 and 1.0 (Figure \ref{fig:caseecc}). Those systems not migrating generally appeared similar to the latter distributions but peaked at a lower maximum eccentricity due to our choice of initial conditions. Finally, some systems had their minimum eccentricity increase, leading to small oscillation magnitudes (Figure \ref{fig:smallosc}). In a small minority of simulations, the maximum eccentricity deviated significantly from Equation \ref{eq:emax}, due to the effect of octupole terms or, for short-period planets,  tides.

%\subsubsection{Known multiplanet systems look like some of these}
%Some of these systems are similar to those discussed in Dawson and Chiang 2014 (citation). [Those that were not in a resonance or coplanar, and also those that were in a heirarchical system.] For those systems we chose a range of inclinations for each of the planets to calculate their true mass, as well as a range of mutual inclinations. From this we were able to determine which systems resulted in a migrating planet due to the eccentric kozai mechanism. Those systems in which the planet exhibited no motion over 3.5 million years (our integration time-scale, chosen simply due to computational constraints) would not be observed in their current configuration and were thrown out. Those systems in which a planet migrated a large distance (>0.1 au) over the duration of the simulation were likewise thrown out, as the likelihood of observing a planet so close to engulfment/circularization is low. 

\subsubsection{Planetary mass effects}\label{sec:masseffect}
% does the mass dependence match the predicted M^2 relationship? The minimum detectable migration needs to be mentioned
The relationship between mass and migration rate, with more massive planets migrating slower and less massive planets migrating rapidly, showed up in all periods (Figure \ref{fig:peridadt}) as a result of Equation \ref{eq:tfriction}. The strength of tides depends on planetary mass and radius, with more massive planets having stronger surface gravity and correspondingly weaker tides. Massive planets do produce larger tides in their host star, but our simulations ignored stellar tides due to their relative weakness, even with large planetary mass. As a result of keeping a constant perturber and planetary radius, small planets migrated the fastest and larger planets the slowest. %Furthermore, the low mass planets run the risk of being disrupted during close approach due to the Roche limit.  

\subsection{Varying system parameters}
\subsubsection{Changing the viscous time-scale}
\begin{figure}
\begin{center}
\includegraphics[width=1.0\columnwidth]{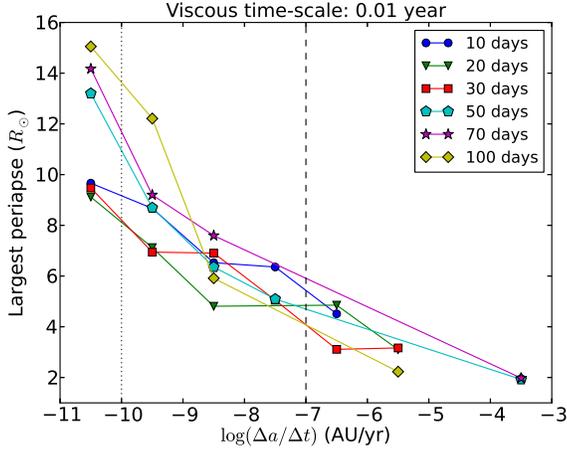}
\caption{Same as Figure \ref{fig:largestperi}, but in the case of a much smaller (0.01) viscous time-scale. Note the larger scale on the y-axis due to smaller maximum eccentricity at similar migration rates. \label{fig:largestperitV0.01}}
\end{center}
\end{figure}

%As shown in Figures so-and-so, 
A viscous time-scale of 0.01 year resulted in planets with higher migration rates and smaller maximum eccentricities than our primary ($t_V=1$ year) simulations (Figure \ref{fig:largestperitV0.01}). Planets on 10-day periods did require  lower eccentricity to migrate inward, and reached much smaller maximum eccentricities than those with longer periods due to their rapid circularization. This shorter viscous time-scale led to fewer planets reaching very high eccentricities, which is more similar to what is seen in observations (Figure \ref{fig:obsWJeccdist}). However, the smaller maximum eccentricities result in larger periapse distances, which results in a larger population surviving to larger stellar radii. Comparison to observations will be discussed more thoroughly in Section \ref{sec:observed}.

%Comparision to Brad's work, Q value from Socrates.

\subsubsection{Changing the perturber}\label{sec:varyperturber}
\begin{figure}
\begin{center}
\includegraphics[width=1.0\columnwidth]{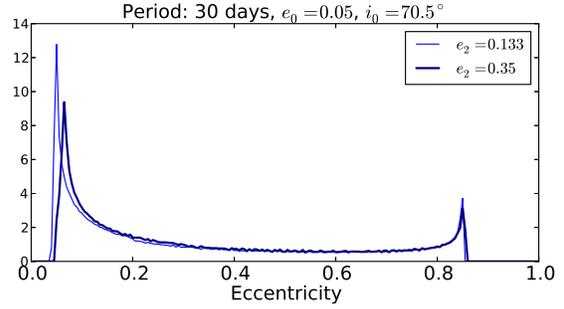}
\caption{Eccentricity frequency distribution of a 30-day period planet (in arbitrary units), for both the default perturber eccentricity (0.133) and the larger value (0.35). The more eccentric perturber produces marginally higher eccentricities on average.\label{fig:e2ecchist}}
\end{center}
\end{figure}

As described in Section \ref{sec:perturberprops}, our simulations also included two smaller samples with altered perturbers for comparison. In the first of these, we increased the perturber eccentricity to near the limit of stability, 0.35, while leaving the other properties (period and mass) the same. The increase in eccentricity resulted in a small shift to larger maximum eccentricities and a resulting slight increase in the overall inward migration rate. The effect was extremely minor, as seen in Figure \ref{fig:e2ecchist}.

The second sample had a dramatically different perturber, one on a significantly larger orbit (10 au $= 10^4$ day period), with a larger mass (30 $M_J$), and greater eccentricity ($e_2=0.64$). The eccentricity value was chosen so that it was also near the limit of stability for the system. For some systems, the larger period resulted in weaker or nonexistent oscillations compared to the closer perturber. This was the case in many of short-period systems through 30 days; the larger period systems averaged more moderate migration rates. Fewer had $da/dt=0$ than with the  close in perturber, due to larger maximum eccentricities, but fewer reached very high migration rates. 

Regardless of perturber or planetary properties, significant $da/dt$ required the planet to reach very large eccentricity values in the vast majority of those with periods longer than 10 days. Additionally, of those undergoing migration without reaching large maximum eccentricity, the majority did so due to a minimum eccentricity above 0.2, a value higher than most observed WJs. 

%While this perturber has some positive effects, it also produces a significantly bigger RV signal
% Both of these objects are going to produce a big RV signal. If anything the more distant perturber would be harder to catch because of the longer time baseline. 

\subsection{Migration rates in the absence of a perturber}

\begin{figure}
\begin{center}
\includegraphics[width=1.0\columnwidth]{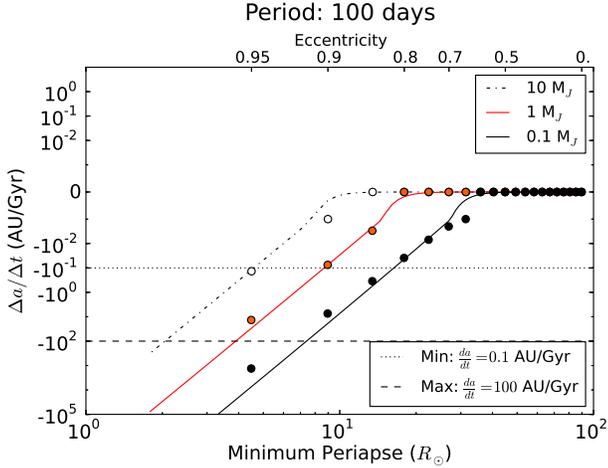}
\caption{Migration rate as a function of periapse distance/eccentricity for 100-day planets without a perturber, and thus non-oscillating eccentricity. The dependence of migration rate on planetary mass and maximum eccentricity is apparent. \label{fig:dadtnoperturb}
}
\end{center}
\end{figure}

 The essence of our model is that currently observed planetary eccentricities substantially underrepresent the rate of tidal
migration experienced by the system because of transient episodes in which the eccentricity oscillates to much larger values.
Thus, as a comparison set we can examine a population of planets whose eccentricities do not oscillate.
These unperturbed planets  behaved as expected, migrating by much larger amounts as compared to oscillating systems of equal maximum eccentricity. As shown by the lines in Figure \ref{fig:dadtnoperturb}, the migration magnitude is well fit by an analytical formula of the form
\begin{align}\label{eq:dadtconst}
\frac{da}{dt}=-4\times10^{-4} f_e(e^2)\left(\frac{a_p}{0.1 \text{ au}}\right)^{-8}\left(\frac{M_p}{1 M_J}\right)^{-2.4}\text{ au/Gyr}
\end{align}
where $f_e(e^2)$ is a function of eccentricity derived from the tidal equations in the case of PS rotation (see Equation \ref{eq:fecc}). Direct calculation of the migration rate leads to a different dependence on SMA and planetary mass (see Equation \ref{eq:psdadt}). The discrepancy is likely due to the planet rotating slightly faster in our simulations. Similar to the oscillating systems, the larger planets migrated less due to experiencing weaker tides from the star, and the larger periods required correspondingly larger eccentricities. 

\section{Stellar evolution effect}\label{sec:rstar}
%Setup for this section
To test if KL oscillations can account for the missing WJs around evolved stars, we must determine the stellar size required to remove each of our simulated planets in the case of both oscillating and constant eccentricity. Here we focus on the planetary systems that could be observed as WJ around other stars. For this reason, we limit the migration rate to $10^{-1}-10^{2}$ au/Gyr as indicated by the lines in Figures \ref{fig:largestperi}, \ref{fig:peridadt}, and \ref{fig:dadtnoperturb}. This rate is rapid enough that WJs can have entered the $10-100$ day-period regime in the lifetime of their star, but long enough that a significant number are observed there. Furthermore, we limit the planet mass to $0.3-3 M_J$. This limit is to avoid being influenced both by both low-mass planets, whose large migration rates may be inaccurate due to our choice of uniform planetary radius, and massive planets, which are more likely to be affected by tides raised on the star that we did not include. 

\begin{figure}
\begin{center}
\includegraphics[width=1.0\columnwidth]{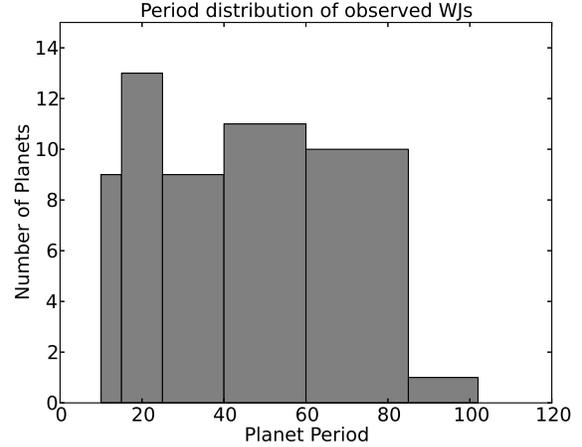}
\caption{The period distribution of observed WJs, binned according to our simulation periods. This distribution informed us how many eccentricity samples to draw from each period.\label{fig:goodperiods}
}
\end{center}
\end{figure}

65 systems across the six orbital period bins meet these criteria. We create a population of planets for comparison by drawing from the period bins according to the observed WJ distribution (Figure  \ref{fig:goodperiods}). With each draw from a given period, we randomly select one of the systems and a value from its eccentricity distribution. We repeat this process until we obtain a final set of 848 planets, each with an eccentricity, period, and planet mass, which match the observed period distribution. These systems represent what the oscillating systems, or an analogous population with constant eccentricity, would look like in observations.  With our population of oscillating- and constant-eccentricity planets, we then determine the criteria for removal.

\subsection{Evolution time-scale}
It is only only appropriate to assume the planet is removed at its maximum eccentricity if the evolution time-scale of the star ($R_*/\dot{R_*}$) is significantly longer than the KL time-scale. In the case of very brief evolution time-scales, the WJ eccentricity would remain relatively constant and the planets would be removed at whatever eccentricity they happened to have at that point in stellar evolution. Using MESA \citep{Paxton:2011ys, Paxton:2013qy} models, we calculated the expansion time-scale of the host star to be $>10^7$ years through $R_*=40R_\odot$, or roughly half the size of our largest planetary orbits. The KL time-scale for our simulations ranged from $7\times10^3 - 8\times10^4$ years, depending on the period of the planet, which are orders of magnitude shorter than the stellar evolution time. As a result, we safely assume the maximum eccentricity determines when the planet is removed via contact. In addition, this portion of stellar evolution occurs without any measurable change in mass, so we can safely ignore the effect of mass loss on the planetary orbits. %Although, see an \emph{amazing} paper by Frewen and Hansen 2014 for some info on that!

\subsection{Planetary removal mechanism}\label{sec:planetremoval}
A WJ can be removed in one of two ways.

 The first is if the planet comes into direct contact with the stellar atmosphere. Hydrodynamic drag during a single
periastron passage will remove binding energy of the order of
\begin{equation}
\Delta E \sim 4 \times 10^{46} ergs \, \rho \left( \frac{R_p}{1 R_J} \right)^2 \frac{M_*}{1 M_{\odot}} 
\end{equation}
where we have assumed a periastron distance $\sim$ stellar radius, $R_p$ is the planet radius and $\rho$ is the density of
the star at the radius of interaction. This is large compared to the orbital binding energy $\sim 2 \times 10^{43} ergs$ of
a Jupiter-mass planet with an orbital period of 100~days, as long as $\rho > 10^{-3} g.cm^{-3}$. This condition is 
satisfied very close to the surface of moderately evolved stars of a few solar radii extent, and so the hydrodynamic drag-down of a planet occurs
on a few orbital timescales once the planet starts to impact the stellar atmosphere.

 A second removal mechanism occurs if the planet can tidally migrate interior to 10 days, becoming a HJ until it undergoes direct contact. This migration occurs more quickly as the star evolves than during the main sequence because the star expands to the point where stellar tides dominate the tidal decay \citep{Villaver:2014gf}. Once this occurs, inward migration increases dramatically with continued stellar expansion due to the $R^8$ dependence in the tidal friction time-scale (Equation \ref{eq:tfriction}). A planet will migrate out of the WJ period space on a time-scale of roughly $P_T = -a/(da/dt)$. 

The value of $da/dt$ caused by tides in the planet, paired with some assumptions about stellar and planetary tides, allow us to quantify the increase in migration rate due to stellar expansion. We take the contribution from the star to be 
\begin{align}\label{eq:Pt}
P_T = \frac{P_{T_p}}{1+(R_*/R_{*,eq})^8}
\end{align}
where  $P_{T_p}$ is the migration time-scale before stellar evolution (due only to planetary tides) and $R_{*,eq}$ is the size of the star at which stellar tides match planetary tides. The value of $P_{T_p}$ for oscillating systems comes from our simulation results, while $P_{T_p}$ for constant eccentricity systems comes from Equation \ref{eq:dadtconst}. Importantly, $f(0)=10^{-3}$ in Equation  \ref{eq:dadtconst}, so that even planets on circular orbits are migrating slowly inward. 

To determine the value of  $R_{*,eq}$, we set the tidal time-scales of the star equal to that of the planet times a coefficient, which accounts for the different spins between the two: $t_{\text{F}_*}=f_s t_{\text{F}_p}$. We assume viscous time-scales of $t_{V_*}=50$ years based the planet-to-star strength from \citet{Hansen:2010lq}, and $f_s=0.2$ based on calculations of $f(e^2, \Omega)$, our function $f(e^2)$ with a non-PS spin value.
\begin{align}
R_{*,eq} = R_p\left(f_s\frac{t_{V_*}}{t_{V_p}} \right)^{1/8} \left(\frac{M_*}{M_p}\right)^{3/8}
\end{align}
We note that this equation assumes the viscous time-scale for the star stays constant over stellar evolution, which is not strictly true. However, the very weak dependence on $t_{V_*}$ means it should not have a significant effect. Using the calculation from \citet{Zahn:1977lr}, $t_{V_*}\propto (L/R^2)^{-4/3}\propto T_\text{eff}^{-4/3}$. For our stellar model, the surface temperature drops from 6300K to 3200K as the star grows to $50R_\odot$, corresponding to an increase in viscous time-scale by a factor of 2.5, or a 12\% increase in $R_{*,eq}$ at its largest. 

We consider a planet with a migration time-scale $P_T< P_\text{short}=10^6$ years to be removed, due to the comparatively brief period of stellar evolution relative to the main sequence lifetime.  Rearranging Equation \ref{eq:Pt}, we can solve for $R_\text{short}$ as a function of migration time-scales and $R_{*,eq}$:
\begin{align}
R_\text{short} = R_{*,eq} \left( \frac{P_{T_p}}{P_\text{rem}} - 1\right)^{1/8}
\end{align}
Therefore we define a planet to be removed when the size of the star reaches $R_\text{rem}$, the smaller of $a(1-e_\text{max})$ and $R_\text{short}$.

\subsection{Stellar expansion results}
% Return to case study
As an example, we draw 10 random samples from the eccentricity distribution of our case study (Figure \ref{fig:caseecc}) and calculate the size of the star when the planet is removed assuming the eccentricity is constant. Table \ref{tab:casestudy} lists these values along with the periapsides and migration rate. The oscillating eccentricity is listed for comparison, showing that only in that case is the planet removed via collision; the constant-eccentricity planets are all removed via migration out of the WJ region, caused by stellar tides.

\begin{table}
\centering
\begin{tabular}{ r r r r}
$e$ & $a(1-e)$ & $da/dt$ & $R_\text{rem}$\\
&$(R_\odot)$& (au/Gyr) &$ (R_\odot)$\\
\hline
$0.07-0.93$& $53-4.0$&-3.0e0&4.0\\
0.11 & 50 & -5.4e-8 & 38 \\
0.22 & 44 & -1.4e-7 & 33 \\
0.25 & 42 & -1.9e-7 & 32 \\
0.46 & 30 & -4.2e-6 & 22 \\
0.15 & 48 & -7.0e-8 & 37 \\
0.67 & 19 & -3.6e-5 & 13 \\
0.25 & 42 & -2.0e-7 & 32 \\
0.56 & 25 & -2.4e-5 & 18 \\
0.91 & 5.0 & -4.1e1 & 2.9 \\
0.46 & 30 & -4.1e-6 & 22 
\end{tabular}
\caption{Migration values and stellar size at planetary engulfment ($R_\text{rem}$) for a simulated oscillating planet and 10 constant-eccentricity realisations.  The $da/dt$ values for the latter group were calculated using Equation \ref{eq:dadtconst}.}
\label{tab:casestudy}
\end{table}

%[ 0.109406,  0.222591,  0.249645,  0.464238,  0.146829,  0.668945, 0.250789,  0.555897,  0.911689,  0.461746]

\begin{figure}
\begin{center}
\includegraphics[width=1.0\columnwidth]{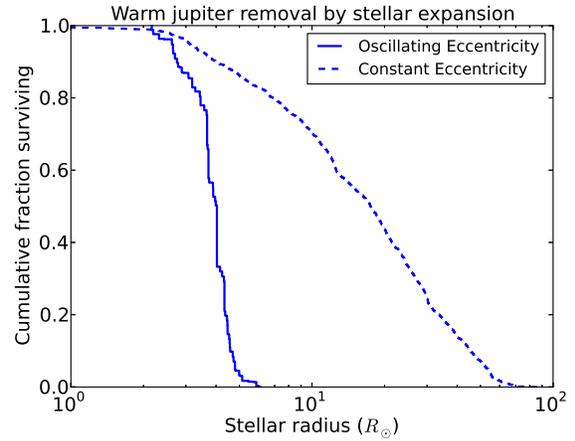}
\caption{The fraction of oscillating and constant-eccentricity WJs that survive as a function of stellar radius. While the fraction of oscillating planets drops off dramatically above $3R_\odot$, the fraction with constant eccentricity is significant even as the stellar radius exceeds $20R_\odot$, which indicates that the lack of WJs around evolved stars can be effectively explained by eccentricity oscillations.\label{fig:cumfrac}}
\end{center}
\end{figure}

We perform the same analysis on all eligible systems, scaling the contribution from systems of each period according to the observed period distribution. Our results are shown in Figure \ref{fig:cumfrac}: the oscillating population drops off as soon as the star exceeds $3R_\odot$, removing all but a handful of planets by $5R_\odot$. By requiring these planets to have a measurable migration rate ($10^{-1}-10^{2}$ au/Gyr), we required them to have small periapsides as well. As a result, those systems are removed almost exclusively by collision with the star. The constant-eccentricity population drops off much more slowly, with some planets surviving until the star is over $50R_\odot$. A small fraction of these planets are on very eccentric orbits  to match the high end of the distribution of oscillating eccentricities. As a result, those planets are removed by collision with the star. In general, however, most had low or moderate eccentricity (as seen in Figure \ref{fig:groupeccdist}) and are removed when the star dominates their migration rate. We note that varying $t_{V*}$ does have an effect for those systems, but only serves to shift the constant-eccentricity population to stellar sizes larger by a factor of $2-3$.

The discrete periods of our simulations are identifiable in the constant eccentricity population as small, steep drops at specific stellar radii. This effect results from the minimum $da/dt$ at a given period: all low-eccentricity planets of a given period have similar $da/dt$ values and are removed at similar stellar radii. When the minimum migration rate is removed, the bumps are smoothed out. The 100-day population between 50 and 60 $R_\odot$ is negligible, due to the small number of such planets in the observed WJ period distribution. 

%In this comparison we have ignored other effects which may serve to excite eccentricity, such as chaotic diffusion. Such effects appear to be capable of producing significant eccentricity values, at least in the context of the solar system \citep{Laskar:2008lq}. However, the 

\section{Comparison to observations}\label{sec:observed}
The model described here is motivated by the claimed deficit of WJs around moderately evolved stars \citep{Johnson:2007kx, Johnson:2011lr}, as seen in Figure \ref{fig:PvRstar} and Section \ref{sec:wjest}. We postulate that the reason for this deficit is that the observed eccentricity distribution of WJs around main sequence stars is really a snapshot of a population whose eccentricities are oscillating via the KL mechanism while they migrate inwards due to tidal friction. The fact that the oscillation time-scale is short compared to the characteristic time-scale for the stellar evolution means that planets are removed from the observed sample when their periapsides oscillate to the minimum value and interact with the host star. Figure 15 shows the result of such a model and demonstrates that, under these conditions, a pre-existing WJ population will be largely removed by the time the stars evolve to 4 $R_\odot$, in contrast to the case where the eccentricities of the observed population do not oscillate. The exact location of WJ removal depends on the details of tidal forces and the perturber, but the general behavior is well described by Figure \ref{fig:cumfrac}.

\begin{figure}
\begin{center}
\includegraphics[width=1.0\columnwidth]{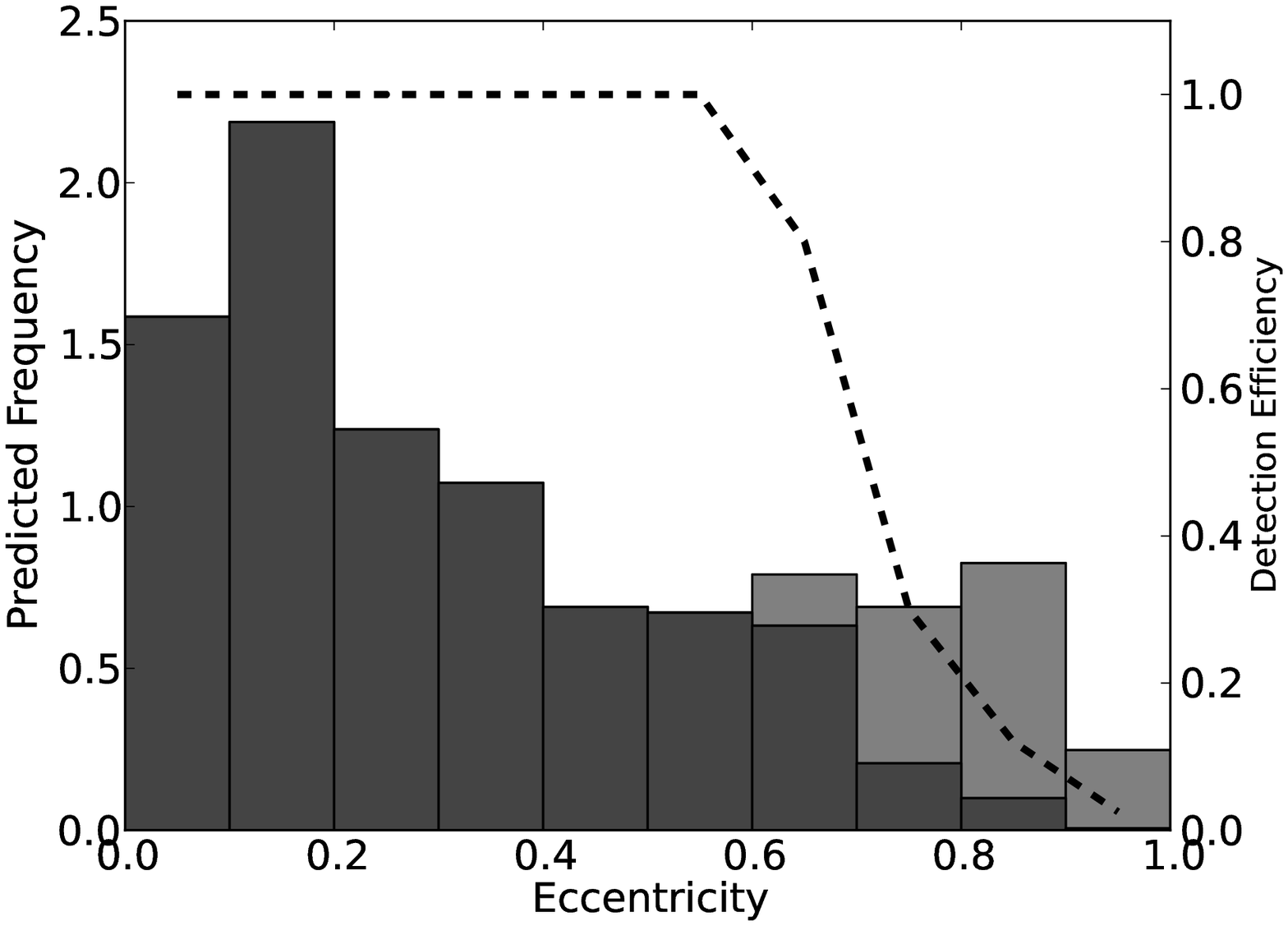}
\caption{The distribution of eccentricity values drawn from our simulated systems, with the same period distribution as observed (grey). The detection efficiency of 100-day planets with signal-to-noise of 10 (dashed line), obtained from \citet{Cumming:2004kx}, drops off dramatically at high eccentricities and produces the eccentricity distribution predicted in observations (dark grey). \label{fig:groupeccdist}}

\includegraphics[width=1.0\columnwidth]{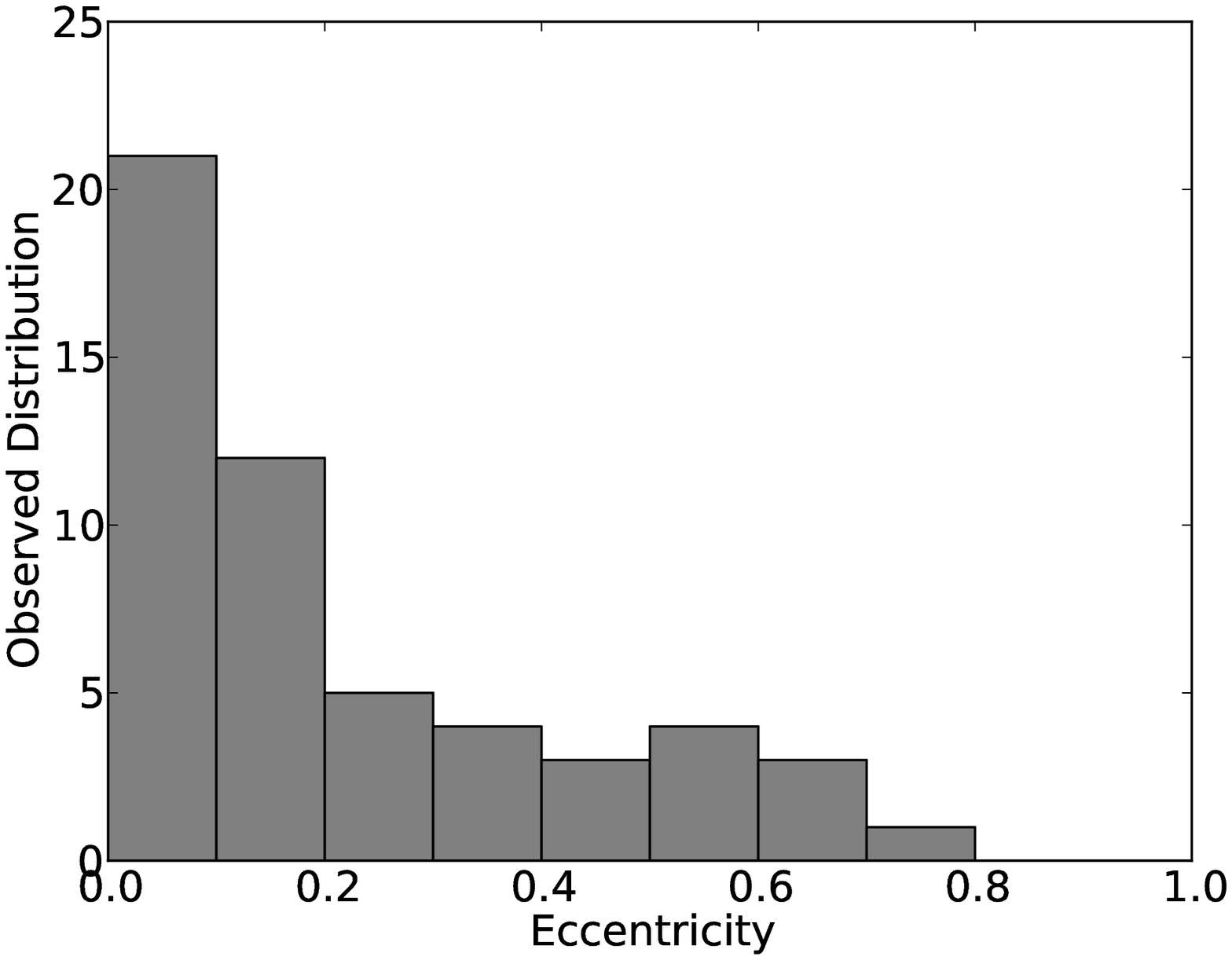}
\caption{The eccentricity distribution for observed WJs, taken from the Exoplanet Orbit Database. The small number of highly eccentric planets differs significantly from the oscillating distribution, but that may be a result of low detection efficiency..\label{fig:obsWJeccdist}}
\end{center}
\end{figure}

However, our results do not match all observations. Figures \ref{fig:groupeccdist} and  \ref{fig:obsWJeccdist} show the distribution of eccentricities for our simulated systems and observed WJs (from the Exoplanet Orbit Database), respectively, with our simulated population drawn from the same period distribution. In both cases the systems are restricted to the Jupiter mass range (0.3-3) and the period range of $10-100$ days. Comparison of these two populations using the Kolmogorov--Smirnov (KS) test gives a $p$-value of $1\times10^{-4}$, indicating they are unlikely to be drawn from the same underlying population. The discrepancy  is primarily due to the significant fraction (15\%) of simulated WJs with high eccentricity ($e>0.8$), while no observed WJs have such high values. The lack of high-eccentricity WJs has also been noted by \citet{Dawson:2015kl}. 

The observed population also includes an excess number of low-eccentricity planets, which is difficult to reconcile with the orbital behavior of our planets. Many of the simulated planets, even those started with eccentricity of 0.05, had a minimum eccentricity peak above 0.1, similar to the eccentricity distribution shown in Figure \ref{fig:caseecc}. This figure also shows the cumulative distribution of eccentricity values; blue, green, and red dashed lines indicate the cumulative observed distribution at 75, 90, and 98 percent of WJs, respectively, for the simulated planet (dotted lines) and eccentricity distribution of all observed WJs (dashed lines). While tidal effects can cause circularisation, the accompanying orbital decay produces HJs, not WJs, on circular orbits.  % (socrates?, dawson and chiang, petrovich?)

\subsection{Observational biases}

 The discrepancy at high eccentricities is a direct consequence of our underlying model.
 In order for host stars to remove their orbiting WJs early on in stellar evolution, as observations imply, the minimum periapsides must be quite small. As a result planets must undergo KL oscillations to large eccentricities, leading to a small but significant fraction of WJs inhabiting that portion of the eccentricity distribution at any given time. Even oscillating systems peaking strongly near $e=0$ have a significant tail at high eccentricities, in conflict with observations. 

However, if eccentric planets are more difficult to detect than low-eccentricity or circular planets, then the dearth of high-eccentricity systems could be an observational effect, not a physical one. Studies of exoplanet detectability in radial velocity surveys \citep{Cumming:2004kx, OToole:2009fj} have shown that that appears to be the case above eccentricities $\sim0.5$, where the largest difference between observed and simulated populations exists. To test how much this effect can improve the fit between our results and observations, we apply the detection efficiency (DE) of Figure 4 in \citet{Cumming:2004kx} to our eccentricity distribution. We use the DE found for fitting to a Lomb-Scargle periodogram with $N=39$ observations short-period (100 day) planets, with a signal-to-noise ratio of 10, shown in Figure \ref{fig:groupeccdist}. After application, the KS test gives a $p$-value of $4\times10^{-4}$, somewhat improved compared to the distribution without correcting for DE.  The remaining mismatch is now a consequence of the excess of
low eccentricities. This can be demonstrated by assuming that the observations contain 10 percent of the population in circular planets,  leading to a $p$-value of 0.03.  Thus, with the correction for DE, a population primarily oscillating is consistent with observations. A population of circular WJs this small would not have a high probability of being detected around evolved stars even if they existed, and could have originated via an alternative migration mechanism, such as disc migration.

Given the specificity of this detection efficiency function and the inclusion of a separate, distinct population, we cannot claim that this calculation proves that our population matches that of observations. An in-depth examination of the detection efficiency of WJs around evolved stars is outside the scope of our work. However, this calculation does show   the conditions required to satisfy observations in such a manner that the KL migration offers a plausible physical explanation for the rapid removal of WJ by stellar evolution. As more eccentricities are determined in systems detected via the transit method, these biases may be reduced, allowing us a better view of the underlying eccentricity distribution. We finish by noting that our DE-corrected distribution predicts that $\sim1$ percent of observed WJs should have eccentricities greater than 0.8. Given that there are currently only 63 WJs listed in the Exoplanet Orbit Database, it is unsurprising that none has high eccentricity. As the number of confirmed WJs increases and improved methods of analysing data are implemented (see \citealt{OToole:2009fj}), the high eccentricity population,  if it exists, should become apparent. 

% Expect to see some when statistics get good enough.
% 

%Another possible observational bias is against detecting planets around evolved stars. It is a significant bias in the case of planetary transits, where the transit signal is diluted due to the greater surface are, but not true in the case of RV surveys, which which are more likely to detect planets around the more slowly rotating evolved  stars. Our comparison focused on planets detected via RV measurements, so we do not think our observational sample was affected.
 
\subsection{Assumptions of physical effects}
Our simulations ignored the effect of stellar tides, which were only included in the evolved star calculations. Larger stellar tides would increase the tidal decay for a planet with a smaller maximum eccentricity and strengthen that decay as the star evolved. However, limits can be placed on the strength of tides in stars from the population of WJs \citep{Hansen:2012ly}. Stellar tides must be weak enough that planets can exist on orbits shorter than one day for an observationally significant amount of time. For that reason, it is unlikely that stellar tides can dominate the evolution of most planets except for the most massive ones. As a test, we simulated 32 systems at 50 days with identical properties to our primary simulations, but with the stellar tidal time-scale set to 50 years. The simulations were qualitatively identical to those without stellar tides, indicating they would need to be significantly stronger than the current limits in order to account for the lack of observed eccentric WJs. 

Our simulations also assumed the equilibrium model for tides, which is an approximation. Tidal effects may differ significantly, both in the star and in the planet, when they are forced on an eccentric orbit. The existence of HJs would not constrain such effects due to their uniformly near-circular orbits. Additionally, we ignored the size difference between 1 $M_J$ planets and 0.1 $M_J$ planets. Correcting for this would likely reduce the migration rate for low-mass planets, leading to a larger population in our defined migrating region. However, many WJs are Jupiter-mass and above, and the issue of a large periapse preventing prompt removal remains.

\section{Conclusion}\label{sec:conc}
%The evolved planets may in fact be not be massive, and thus a mechanism caused by stellar evolution would be necessary to account for the different population Lloyd:2011dp, Lloyd:2013dq
A number of planets have been found around evolved stars, but there appears to be a lack of massive planets interior to 0.6 au \citep{Johnson:2007kx, Bowler:2010uq, Johnson:2011lr}. Two possibilities exist: either the underlying population of planets differs around the unevolved progenitors of these generally more massive ($>1.5M_\odot$) stars, or stellar evolution has led to their removal. The results of \citet{Lloyd:2011dp, Lloyd:2013dq} have called into question whether the evolved stars truly originate from a more massive population, supporting the latter reason for the absence of WJs. Additionally, \citet{Schlaufman:2013cr} showed that some evolved stars have a different population of planets than their unevolved progenitors of the same mass, indicating that stellar evolution is a cause in at least some cases. Most recently, \citet{Johnson:2014dq} showed that at least one evolved star in the disputed population has a mass truly greater than 1.5 $M_\odot$, as they claimed in prior works. Taken together, these results leave considerable ambiguity for the explanation of missing WJs.
%although \citet{Johnson:2014dq} showed that in at least one case an evolved star is very likely larger than $1.5M_\odot$.

Here we have simulated planets undergoing KL oscillations as part of their migration inward and examined how the population decays with stellar evolution. By using a model population of WJs and their perturbing companions, we have shown that KL oscillating WJs explain the observed absence around evolved stars better than a constant-eccentricity population. A population of migrating, KL oscillating WJs is almost entirely removed around an evolving star by the time it reaches $5R_\odot$, while an observationally identical population with constant eccentricity survives stellar expansion beyond $40R_\odot$. Finally, although we have adopted a stellar mass of 1.2 $M_\odot$ in our simulations, it should be noted that the rapid removal of WJs migrating via KL oscillations is applicable regardless of stellar mass. Therefore the absence observed by  \citet{Johnson:2007kx} and related works need not indicate that WJs are absent around more massive stars in general.

\section*{Acknowledgments}
This research is supported by the Dissertation Year Fellowship at UCLA. We thank Smadar Naoz for the use of her code and helpful comments, as well Jean-Luc Margot for his comments.

\appendix
\section{Planetary migration during pseudo-synchronous rotation}
\noindent The orbital evolution of a planet due to tides is given by
\begin{align}
\frac{da/dt}{a} &= -2\left[W_p + W_* + \frac{e^2}{1-e^2}(V_p + V_*)\right]
\end{align}
where the subscripts $p$ and $*$ correspond to the planet and host star, respectively. $V$ and $W$ are given in \citet{Eggleton:2001lr}:
\begin{align}
V &= \frac{9}{t_{F}}\bigg\{\frac{1 +(15/4)e^2 + (15/8)e^4 + (5/64)e^6}{(1-e^2)^{13/2}} \\ \nonumber
    &-\frac{11\Omega}{18n}\frac{1+(3/2)e^2+(1/8)e^4}{(1-e^2)^5}\bigg\}\\
W &= \frac{1}{t_{F}}\bigg\{\frac{1 +(15/2)e^2 + (45/8)e^4 + (5/16)e^6}{(1-e^2)^{13/2}} \\ \nonumber
     &-\frac{\Omega}{n}\frac{1+3e^2+(3/8)e^4}{(1-e^2)^5} \bigg\}
\end{align}
where $n$ is the mean motion of the orbit and $\Omega$ is the rotation rate of the body. A migrating WJ will have already reached pseudo-synchronous rotation, which occurs when $W_p=0$ \citep{Hut:1981pd}. The rotation rate in that case is given by
\begin{align}
\frac{\Omega_{ps}}{n} = \frac{1+(15/2)e^2+(45/8)e^4 + (5/16)e^6}{(1+3e^2+(3/8)e^4)(1-e^2)^{3/2)}}
\end{align}
Plugging in $\Omega_{ps}$, we get the strength of tides for a pseudo-synchronous planet:
\begin{align}
V_p(\Omega_{ps}) &= \frac{9}{t_{Fp}} \bigg\{1792+5760e^2+14336e^4 \\ \nonumber 
 &+ 5480e^6+1020e^8+25e^{10}\bigg\}\frac{1}{4608(1-e^2)^{15/2}}
\end{align}
Assuming planetary tides dominate during the main sequence ($V_p>>V_*$) and that the planet is in PS rotation ($W_p=0$),  we can simplify the tidal decay equation:
\begin{align}
\nonumber
\frac{da/dt}{a}&=-2\left[ W_* + \frac{e^2}{1-e^2}(V_p)\right]\\ \nonumber
&=-2W_*  - \frac{2e^2}{1-e^2}\frac{9}{t_{Fp}} \bigg\{1792+5760e^2+14336e^4 \\ \nonumber 
 &+ 5480e^6+1020e^8+25e^{10}\bigg\}\frac{1}{4608(1-e^2)^{15/2}}\\ 
&=-\frac{2f_e(e^2)}{t_{Fp}}
\end{align}
where
\begin{align}\label{eq:fecc}
 \nonumber 
f_e(e^2) &= t_{Fp}W_*  +  \bigg\{1792e^2+5760e^4+14336e^6 \\
 &+ 5480e^8+1020e^{10}+25e^{12}\bigg\}\frac{1}{512(1-e^2)^{17/2}}
\end{align}
We have retained $W_*$ because it tends to $t_{F*}^{-1}$ as $e\rightarrow 0$, assuming the star is rotating slowly, while the remainder of the expression tends to 0. The limits for $f_e(e^2)$ are therefore $t_{Fp}/t_{F*}$ near $e=0$ and $3.5(1-e^2)^{17/2}$ for $e\sim1$. From the definition of $t_F$, Equation \ref{eq:tfriction}:
\begin{align}
\frac{t_{Fp}}{t_{F*}} =\frac{t_{Vp}}{t_{V*}}\left(\frac{R_p}{R_*}\right)^{-8}\left(\frac{M_p}{M_*}\right)^{3}\left(\frac{1+2k_p}{1+2k_*}\right)^{-2}
\end{align}
In Section \ref{sec:tidaldecay} we assume the following values for the planetary systems: $t_{Vp}=1$ year, $t_{V*}=50$ years, $M_*=1.2M_\odot$, $M_p=0.1-10M_J$, $k_p=0.25$, and $k_*=0.014$. Entering these values, we we get a ratio ranging from $10^{-6}-1$. For the subset in Sections \ref{sec:rstar} and \ref{sec:observed}, $0.3-3M_J$, the values range from $2.7\times10^{-5}-2.6\times10^{-2}$. We assume $t_{Fp}/t_{F*}=10^{-3}$ in all cases for simplicity, and note that it is a small correction in all cases. 

Plugging in for $t_{Fp}$ using Equation \ref{eq:tfriction}, we get the expected value for tidal migration:
\begin{align}\label{eq:psdadt}
\frac{da}{dt} = -18f_e(e^2)\frac{a}{t_{Vp}}\left(\frac{R_p}{a}\right)^8 \left(\frac{M_*}{M_p}\right)^2\frac{1}{(1+2k_p)^2}
\end{align}

\bibliographystyle{mn2e}   
\bibliography{/Users/sfrewen/Research/research}

\label{lastpage}

\end{document}